\documentclass[structabstract]{aa}  
\usepackage{longtable}
\usepackage{rotating}
\usepackage{lscape}
\usepackage{latexsym}
\usepackage{natbib}
\usepackage{graphicx}
\usepackage{color}
\usepackage{ulem}
\usepackage{txfonts}
\begin{document}
   \title{Near-infrared photometry of Galactic planetary nebulae with the VVV Survey}
%Accurate photometry of planetary nebulae with VVV}

%   \subtitle{I. Overviewing the $\kappa$-mechanism}

   \author{W. A. Weidmann
          \inst{1}\fnmsep\thanks{Member of Carrera del Investigador CONICET, Argentina.},
          R. Gamen\inst{2}\fnmsep\thanks{Member of Carrera del Investigador CONICET, Argentina.},   
P.A.M. van Hoof\inst{3}, %\fnmsep\thanks{}  
A. Zijlstra\inst{4},    
D. Minniti\inst{5}\fnmsep\inst{6}\fnmsep\inst{7}
  \and 
M. G. Volpe\inst{8}
}

   \institute{Observatorio Astron\'omico C\'ordoba, Universidad Nacional de C\'ordoba, Argentina
              \\
              \email{walter@mail.oac.uncor.edu}
         \and
             Instituto de Astrof\'isica de La Plata, CCT La Plata-CONICET, Universidad Nacional de La Plata, Argentina\\
             \email{rgamen@fcaglp.unlp.edu.ar}
%             \thanks{The university of heaven temporarily does not accept e-mails}
      %
           \and
             Royal Observatory of Belgium, Ringlaan 3, 1180 Brussels, Belgium
    \and
    Jodrell Bank Centre for Astrophysics, University of Manchester, Manchester M13 9PL, UK
   \and      
        Departamento de Astronom\'ia y Astrof\'isica, Pontificia Universidad Cat\'olica de Chile, Av. Vicu\~na Mackenna 4860, Casilla 306, Santiago 22, Chile \\  
             \email{dante@astro.puc.cl}
\and
Vatican Observatory, V00120 Vatican City State, Italy
  \and 
Departamento de Ciencia Fisicas, Universidad Andres Bello, Santiago, Chile
\and
Instituto de Astronom\'ia Te\'orica y Experimental, Argentina
}

%   \date{Received September 15, 1996; accepted March 16, 1997}

% \abstract{}{}{}{}{} 
% 5 {} token are mandatory
 
  \abstract
  {Planetary nebulae (PNe) are powerful
  tracers of evolved stellar populations. Among the 3000 known PNe in the
  Galaxy, about 600 are located within the 520 square-degree area covered by
  the VVV survey. The VVV photometric catalogue provides an important new
  dataset for the study of PNe, with high-resolution imaging in five
  near-infrared bands.  }
   {There are various colour-colour diagrams that can be obtained from the
VVV filters. We investigate the location of PNe in these diagrams and the
separation from other types of objects. This includes the new \textit{Y$-$J} vs. \textit{Z$-$Y}
diagram. }
    {Aperture photometry of known PNe in the VVV area was
   retrieved from source catalogues. Care was taken
   to minimise any confusion with field stars. The colours of the PNe we are
   determined for (\textit{H$-$K$_{\rm s}$}), (\textit{J$-$H}), (\textit{Z$-$Y}), and 
  (\textit{Y$-$J}), and compared
   to stars and to other types of emission line objects. Cloudy
   photo-ionisation models were used to predict colours for typical PNe. }
  {We present near-infrared photometry for 353 known PNe.  The
    best separation from other objects is obtained in the (\textit{H$-$K$_{\rm s}$}) vs.
    (\textit{J$-$H}) diagram. We calculated the emission-line contribution to the
    in-band flux based on a model for NGC 6720: we find that this is highest
    in the Z and Y bands at over 50\%, lower in the J band at 40\%, and
    lowest in the H and Ks bands at 20\%.  A new view of PNe in the wavelength
    domain of the $\rm Z$ and $\rm Y$ bands is shown. Photo-ionisation models
    are used to explore the observed colours in these bands. The Y band is
    shown to be dominated by He\,{\sc i} 1.083$\mu$m and He\,{\sc ii} 
 1.012$\mu$m, and
  colours involving this band are very sensitive to the temperature of the
  ionizing star.  }
   {The VVV survey represents a unique dataset for studing crowded and obscured
     regions in the Galactic plane. The diagnostic diagrams presented here
     allow one to study the properties of known PNe and to uncover objects not
     previously classified. 
}

%%%%%%%%%%%%%%%%%%%%%%%%%%%%%%%%%%%%%%%%%%%
%%%%%%%%%%%%%%%%%%%%%%%%%%%%%%%%%%%%%%%%%%%
%%%%%%%%%%%%%%%%%%%%%%%%%%%%%%%%%%%%%%%%%%%

\titlerunning{Planetary Nebulae with the VVV Survey}
\authorrunning{Weidmann et al.}

   \maketitle
%
%________________________________________________________________

\section{Introduction}
%%%%%%%%%5
%PN
%%%%%%%%%%%

Planetary nebulae (PNe) are luminous and short-lived products of evolved
low-mass stars ($0.8 M_{\odot}<M<8 M_{\odot}$). The stars eject much of their
envelopes during a phase of strong mass loss at the end of the asymptotic giant
branch (AGB). The ejecta become ionised after a brief transition phase, when
the star rapidly increases in temperature, the ionised ejecta form a visible
 planetary nebula.  The ejected material is enriched in a variety of
elements (He, C, N, O, Ne, Mg, s-process elements), through nuclear burning and
dredge-up during the AGB phase. The transfer of this gas to the interstellar
medium affects the overall chemical evolution of the Galaxy.  PNe also provide
information about the recent star-forming history in our Galaxy
\citep[e.g.][]{2003IAUS..209..551M}.  Thus, PNe are key objects in the study
of stellar and galactic evolution.

The distribution of known Galactic PNe shows a lack of objects within the
Galactic plane and towards the central regions of the Galaxy
\citep[e.g.][Fig.~6c]{2008MNRAS.384..525M}. This is caused by  high
extinction and by severe crowding. To obtain a better understanding
of the whole Galactic PN population, deeper surveys are needed in these
regions.  Near-infrared (NIR) observations provide an important tool, because they
are less affected by extinction and benefit from better seeing.

The NIR fluxes of PNe depend on a broad range of emission mechanisms,
including warm dust continuum emission, ionic and atomic permitted and
forbidden line transitions, and the underlying free-free and bound-free
components of gaseous continuum emission. Continuum emission from cool stellar
companions, or the PN central star itself, may also contribute.  Broad band IR
photometry of galactic PNe has been reported in the literature by several
authors
\citep[e.g.][]{1973MNRAS.161..145A,1985MNRAS.213...59W,1987RMxAA..14..534P,1997A&AS..126..479G}.
The complete and homogeneous 2MASS survey has substantially increased the number
of PNe with NIR observations \citep{2005MNRAS.357..732R,2009MNRAS.394.1875P},
but was limited by its 4-arcsec spatial resolution.  The VISTA Variables in
the Via Lactea (VVV\footnote{http://vvvsurvey.org/}) public survey now
provides deeper and higher resolution observational data, and will greatly
improve the number of PNe observations in extincted and/or crowded regions of
the sky \citep{2010NewA...15..433M}.

The past studies of PNe have used the $JHK$ bands.  The VVV survey adds two
original bands, the $Z$ and $Y$ bands\footnote{We found only a single paper
  using the $Y$ broad band \citep{2011A&A...531A.157M}.}.  Therefore, our
motivations are to improve the infrared photometry of PNe in order to
contribute to the knowledge of stellar evolution, and to generate tools for
characterising the NIR flux of PNe.

%%%%%%%%%%%%%%%%%%%%%%%%%%%%%%%%%%%%%%%%%%%%%%%%%%%%%%%
%%%%%%%%%%%%%%%%%%%%%%%%%%%%%%%%%%%%%%%%%%%%%%%%%%%%%%%
\section{The VVV observational database}\label{dbase}
%%%%%%%%%%%%%%%%%%%%%%%%%%%%%%%%%%%%%%%%%%%%%%%%%%%%%%%
%%%%%%%%%%%%%%%%%%%%%%%%%%%%%%%%%%%%%%%%%%%%%%%%%%%%%%%

The VISTA 4.1m telescope at the Paranal Observatory
\citep{2010Msngr.139....2E} is equiped with the instrument VIRCAM (VISTA
InfraRed CAMera; \citealt{2006Msngr.126...41E};
\citealt{2006SPIE.6269E..30D}).  The camera contains 16 near infrared
detectors, of $2048\times2048$ pixels each. The pixel size corresponds to 0.34
arcsec on the sky. There are gaps between the detectors so that a single
exposure gives a non-contiguous sky coverage. Such a single exposure is
called a pawprint. By combining 6 \textit{pawprints} with appropriate offsets, a
contiguous coverage of a field is achieved, where all pixels (apart from those
at the edges) have at least two independent exposures.  In the VISTA
terminology such a field is called a \textit{tile}; it covers a 1.64 square
deg field of view, 1.5 square deg of which is covered by at least 2 pawprints.

The data reduction was carried out in the typical manner for infrared imaging.
Details of the procedure are described in \citet{2004SPIE.5493..411I}.  The
median image quality measured on the reduced VVV tile images is around
0.8$\arcsec$ for $K_{\rm s}$, 0.9$\arcsec$ for the $J$ band, and up to
1.0$\arcsec$ for the $Z$-band.

The data calibration is performed by the VISTA Data
Flow System (VDFS) pipeline at the Cambridge Astronomy
Survey Unit (CASU). The photometry is tied to unsaturated
2MASS stars present in the VVV images, even for
the $Z$ and $Y$ filters (not observed by 2MASS), for which
colour equations are applied\footnote{More details about 
VISTA data processing is available from
http://casu.ast.cam.ac.uk/surveys-projects/vista/data-processing}.
For more detail about photometric calibration see
\citet{2011A&A...534A...3G} for all 5 filters.

In this paper we use the magnitudes from the first data release of the VVV
Survey \citep{2012A&A...537A.107S}.  The VISTA Science Archive\footnote{http://horus.roe.ac.uk/vsa}
 provides two different catalogues,
\textit{vvvDetection} and \textit{vvvSource}. The first one contains the
individual detections for sources originating from multiframe images taken
from the VVV. The second one lists the merged sources (in the five bands) from
detections in \textit{vvvDetection} catalogue \citep{2012A&A...537A.107S}.  In
our catalogue, we used the Source-table, with a 2.8$\arcsec$ aperture
diameter.

The catalogue contains a flag to indicate the most probable morphological
classification, in particular 
``-1" is used to denote stellar objects,
``-2" borderline stellar,
``0" is noise, and
``1" is used for non-stellar objects. 
There are also further flags:
``-7", denoting sources containing bad pixels, and the flag
``-9" is used for saturated stars \citep{2004SPIE.5493..411I}.
This makes it possible to reject some bad measurements
and to identify extended emission sources.

Objects with \textit{$J<11.8$}, \textit{$H<11.0$}, or \textit{$K_{\rm s}<11.0$} are
saturated.  We considered \textit{$ZY<$12.5} to indicate saturation, taking into
account that the magnitudes in these two bands are probably 
not as well calibrated as those at $JHK_{\rm s}$  
due to the extrapolation needed.
The magnitudes at which saturation occurs are
dependent on the fields, partly because of seeing, and partly because there
are some differences in exposure times between the bulge
($-10\,^{\circ}<l<10\,^{\circ}$ and $-10\,^{\circ}<b<5\,^{\circ}$) and disc
($-65\,^{\circ}<l<-10\,^{\circ}$ and $-2\,^{\circ}<b<2\,^{\circ}$). We prefer
to adopt the limits shown above.

The magnitudes from the aperture photometry are extracted from the catalogues
following the equation ${\rm mag = zp - (airm - 1 )} \times C - 2.5 \times
\log_{10}(F/T) - {\rm AC}$, where zp= photometric zero-point for default
extinction, airm = air mass, C= default extinction in the passband, F=
aperture flux, T= exposure times and AC= aperture correction in magnitudes.

We refer the reader to~\citet{2012A&A...537A.107S}
for a description of the photometric errors in the
$ZYJHK_{\rm s}$ filters as a function of the magnitude at different
levels of crowding along the VVV area\footnote{
Errors in the fluxes are based on Poissonian statistics. A comprehensive description 
is available at \cite{2012A&A...548A.119C}} 
.

%%%%%%%%%%%%%%%%%%%%%%%%%%%%%%%%%%%%%%%%%%%%%%%%%%%%%%%
%%%%%%%%%%%%%%%%%%%%%%%%%%%%%%%%%%%%%%%%%%%%%%%%%%%%%%%
\section{Sources of 0.8-2.1$\mu$m emission in planetary nebulae in the VVV filters }
%%%%%%%%%%%%%%%%%%%%%%%%%%%%%%%%%%%%%%%%%%%%%%%%%%%%%%%
%%%%%%%%%%%%%%%%%%%%%%%%%%%%%%%%%%%%%%%%%%%%%%%%%%%%%%%

For a better understanding of the results shown in this paper it is worth
studying the characteristics of the near infrared emission observed in PNe.
\citet{1985MNRAS.213...59W} gave a complete description of the principal
emission sources in the J, H, K$_{\rm s}$ bands, thermal plasma continuum,
thermal dust emission, hot stellar continuum, emission lines and cool stellar
continuum from a companion. Their discussion is based on the SAAO filters
which slightly differ from other systems. In particular, the SAAO J-band
includes the strong HeI 1.083 micron line at 50\% transmission, which can
dominate the flux in that band. The VISTA (and 2MASS) J-band filter excludes
this line and it marginally intrudes instead in the Y-filter.

\begin{figure*}
   \centering \includegraphics[width=19cm]{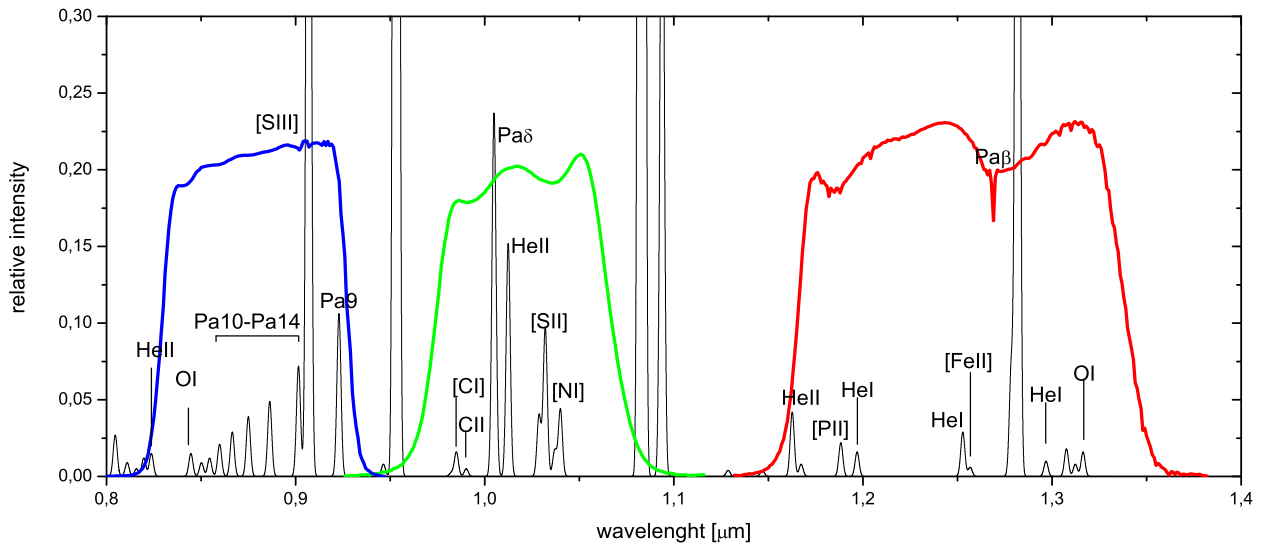}
      \caption{Transmission curves for the $Z$, $Y$ and $J$ broad-band filters
(the effective wavelengths are 0.878, 1.021 and 1.254 $\mu$m respectively),
 the filters are multiplied by a model atmosphere.
The spectrum of the PN IC~5117 is superimposed \citep{2001AJ....121..362R},
scaled to Pa$\beta$=1.}
         \label{filtro}
   \end{figure*}

In the new photometric bands Z and Y, the dust contribution is neglible.
Emission from free-free (FF) and bound-free (BF) transitions is marginally
weaker.  The stellar continuum gains importance, however the major flux
contribution comes from emission lines. The Z band is dominated by
[\ion{S}{iii}] at 0.9069$\mu$m, and the Y band by Pa$ \delta$ and \ion{He}{ii}
at 1.0124$\mu$m.  Figure~\ref{filtro} shows the $ZYJ$ transmission curves, in
comparison to the spectrum of IC~5117, a medium excitation-class planetary
nebula.

The 2MASS J band is somewhat wider than the VISTA J band. But the dominant
Pa$\beta$ is covered by both filters, whilst neither filter contains other
significant emission lines. Therefore, we do not expect a major discrepancy
between both magnitudes.

IPHAS (The INT/WFC Photometric H-alpha Survey, \citealt{2005MNRAS.362..753D}),
its southern extension VPHAS, and DENIS (Deep Near-Infrared Southern Sky
Survey, \citealt{1997Msngr..87...27E}) contain two optical bands that overlap
in some spectral range with the Z VISTA band: these are \textit{i'} (centred
at 0.7743$\mu$m) and I (0.8000$\mu$m) respectively.  
However, magnitudes obtained from
these optical bands cannot be compared with the Z one because the filters are
too different.

To further test the emission mechanisms contributing to the VVV bands, we made
use of the published photo-ionisation model of \citet{2010A&A...518L.137V} for
NGC 6720, the Ring Nebula. This is the archetypal planetary nebula; the nebula
is bipolar, seen almost pole-on, optically thick to ionizing radiation except
towards the poles. The Ring Nebula has a central star which has already
entered the cooling track, and in response to the fading of the star parts of
the outer nebula have begun to recombine. The photo-ionisation model uses as
ionizing source a H-Ni model stellar atmosphere from
\citet{2003A&A...403..709R} with a temperature of $T_{\rm eff} = 135\,$kK. The
electron density of the model is $n_e = 416 \,\rm cm^{-3}$.  For comparison,
IC 5117 shown in Fig.~\ref{filtro} has $T_{\rm eff} = 120\,$kK and density
$n_e = 9\times 10^4\rm cm^{-3}$ \citep{2001ApJ...563..889H}.  The model was
calculated with version C10.00 of the Cloudy photo-ionisation code, last
described by \citet{1998PASP..110..761F}. The model incorporates the neutral
region, to capture molecular hydrogen emission.  We used this Cloudy model to
calculate the free-free and bound-free continuum from both hydrogen and
helium, and lines from ionised and neutral species, and from molecular
hydrogen. The stellar continuum is also included but is not important at these
long wavelengths and high stellar temperatures. The output model spectrum was
convolved with the transmission curves from the VISTA filters combined with the
telescope and instrument throughput and the detector sensitivity. Note that
the model is for a nebula without interstellar reddening but does include a
small amount of internal reddening.

Table~\ref{n6720_lines} shows the detailed list of major lines which
contribute to the in-band flux.  The Z and J-bands are dominated by one
line. The Y band also has one dominant line (He I) but it is 
located at the edge of the transmission curve.  H and K$_{\rm s}$ have 
a scattering of fainter
lines. The total contribution of emission lines to the in-band flux is much
higher for Y and Z than for the other three bands.  The Z and Y-filters have a
strong line close to the edge of the transmission curve, and small changes in
the filter responses can have significant effects on the line contributions.
The specific line contributions will also depend on the excitation of the
nebula especially for the Z-band.

\begin{table}
\caption{Line contribution from the (unreddened) NGC~6720 model results.  The
  columns give the element, wavelength, log flux in usual units, the line
  intensity relative to H$\beta$, and the combined transmission of the sky,
  telescope, filter, and detector at this wavelength. Note that the model
  fluxes have not been multiplied with the transmission. The Z, Y and J bands
  are dominated by one line (but for the Y band it is located at the edge of
  the filter). H and K$_{\rm s}$ have a scattering of fainter lines. The total
  contribution of emission lines to the in-band flux is also given. This is
  much higher for Y and Z than for the other three bands.  }
\label{n6720_lines}      
\centering                          
\begin{tabular}{l c c c c }       
\hline\hline                 
line & $\lambda$ & flux                           & $I$(line)/$I$(H$\beta$) & transm. \\    
     &  [$\mu$m]     & [log erg\,cm$^{-2}$\,s$^{-1}$] &                         & \\
\hline                       
%{\it $Z$ band }\\   
\multicolumn{3}{l}{$Z$ band (0.878$\pm$0.097~$\mu$m)}\\
\multicolumn{3}{l}{line contribution 55\%}\\
{} [S III]{} &  0.9069 &  -10.762 &  0.1591 & 0.717 \\
H  I    &  0.9229 & -11.576 &  0.0244 & 0.644 \\
H  1    &  0.9015 & -11.716 &  0.0177 & 0.712 \\
H  1    &  0.8863 & -11.842 &  0.0132 & 0.707 \\
H  1    &  0.8750 & -11.956 &  0.0102 & 0.698 \\
{} [Cl II] &  0.8579 & -12.012 &  0.0089 & 0.677 \\
 \hline
%{\it $Y$ band }  \\   % (1.021 0.093)
\multicolumn{3}{l}{$Y$ band (1.021$\pm$0.093~$\mu$m)}\\
\multicolumn{3}{l}{line contribution 55\%}\\
He I   & 1.083       &  -10.342  & 0.4182 & 0.048 \\
{} [C  I] &  0.985  &  -11.166  & 0.0627 & 0.597 \\
H I    & 1.005    &  -11.236  & 0.0534 & 0.645 \\
He II  & 1.012    &  -11.305  & 0.0456 & 0.669 \\
{} [S  II] & 1.033  & -11.382  & 0.0382 & 0.640 \\ 
{} [N I] & 1.040   &  -12.026  & 0.0087 & 0.645 \\ 
\hline
%{\it $J$ band } \\  % (1.254 0.172)
\multicolumn{3}{l}{$J$ band (1.254$\pm$0.172~$\mu$m)}\\
\multicolumn{3}{l}{line contribution 39\%}\\
H  I  & 1.282    &  -10.772 & 0.1555 & 0.680 \\
He II & 1.163    &  -12.026 &  0.0087 & 0.226 \\
Fe II & 1.257    &  -12.109 &  0.0072 & 0.730 \\
He I  & 1.278    &  -12.135 &  0.0067 & 0.668 \\
\hline 
%{\it $H$ band } \\  % (1.646 0.291)
\multicolumn{3}{l}{$H$ band (1.646$\pm$0.291~$\mu$m)}\\
\multicolumn{3}{l}{line contribution 21\%}\\
H I &   1.817 & -11.890 &  0.0118 & 0.040 \\
H I &   1.736 & -12.031 &  0.0086 & 0.843 \\
H I &   1.681 & -12.158 &  0.0064 & 0.850 \\
Fe II & 1.644 & -12.182 &  0.0061 & 0.825 \\
\hline 
%{\it $K_{\rm  s}$ band } \\  % (2.149 0.309)
\multicolumn{3}{l}{$K_{\rm  s}$ band (2.149$\pm$0.309~$\mu$m)}\\
\multicolumn{3}{l}{line contribution 22\%}\\
H  I & 2.166   & -11.549  & 0.0260 & 0.833 \\
He I & 2.058   & -11.769  & 0.0156 & 0.616 \\
    \hline                                   
\end{tabular}
\tablefoot{Next to the band are indicated the effective 
wavelengths and the width of VISTA's filters.}
\end{table}

%%%%%%%%%%%%%%%%%%%%%%%%%%%%%%%%%%%%%%%%%%%%%%%%%%%%%%%
%%%%%%%%%%%%%%%%%%%%%%%%%%%%%%%%%%%%%%%%%%%%%%%%%%%%%%%
\section{Results}
%%%%%%%%%%%%%%%%%%%%%%%%%%%%%%%%%%%%%%%%%%%%%%%%%%%%%%%
%%%%%%%%%%%%%%%%%%%%%%%%%%%%%%%%%%%%%%%%%%%%%%%%%%%%%%%

\subsection{Description of the sample}

There are 579 known PNe listed by \citet{1992secg.book.....A}, \citet[MASH
  I]{2006MNRAS.373...79P}, and \citet[MASH II]{2008MNRAS.384..525M} whose
coordinates (Kerber et al., 2003) fall within the VVV area\footnote{ The
IPHAS survey \citep{2009A&A...504..291V} does not overlap with the VVV area.}.

We retrieved the processed and calibrated $ZYJHK_{\rm  s}$ 
images (tiles) from the Cambridge Astronomical Survey
Unit (CASU VIRCAM pipeline v1.1\footnote{Data release 1.1,  
http://casu.ast.cam.ac.uk/surveys-projects/vista}; Irwin et al. 2004)
and also retrieved the MASH H$\alpha$ images \citep{2005MNRAS.362..689P}. 
The planetary nebulae were identified from the H$\alpha$ images using
\cite{2003A&A...408.1029K}, the MASH catalogue and identification charts
\citep{1988AJ.....95..804K,1999ApJ...515..610B,2004A&A...419..563J}.  We
identified the related VVV source taking into account its symmetry with
respect to the H$\alpha$ emission. The visual examination was
essential. Cross-correlation between the VVV and PNe catalogues
(\citealp{2003A&A...408.1029K} for example) did not give satisfactory results
for many sources.  Therefore, the VVV sources were associated with the PNe by
eye.  Up to 82\%\ of the sample are closer than 1.4 arcsec to the catalogue
coordinates of \cite{2003A&A...408.1029K}.

From the initial sample of 579 PNe, 75 are highly extended sources and
these will be considered in future papers,  
and 123 PNe could
not be related to any source in the VVV images,
perhaps because they have low surface brightness or the central star is too
weak in the NIR bands.  
The remaining PNe analysed in this paper are related
to sources in the VVV catalogues with 
NIR emission mainly enclosed in an area of 1.4~arcsec of radius
(even in PNe of high surface brightness and appreciable angular
size); in this sense we are observing the emission from the central regions of
the nebula. The integrated flux may be underestimated where the nebula is
larger than the aperture used here.
See Appendix~\ref{apendi} for a discussion about the used aperture.

In our catalogue we divide the data into five categories 
(see Table~\ref{tablita} and \ref{tablita-2}):

\begin{itemize}
\item Objects with photometric data (not necessarily in the five bands).
\item NIR emission of the PN is detected in tile frames,
but there are no photometric data in the VSA database (S/D).
\item Extended objects, where the NIR emission is larger than $3\arcsec$ in diameter (ext).
\item PNe extended (in H$\alpha$) without NIR sources in the regions of their
  geometrical centre (N/D).
\item Objects with unreliable photometry, i.e. large uncertainty in the
  magnitudes ($>$0.2), 
 classified as noise or saturated by the pipeline (bad ph.). 
\end{itemize}

Nearly 80\% of the sample is detected in at least one band.  Although the
integration times in the region of the bulge are shorter than in the plane (4s
and 10s per frame for the $K_{\rm s}$ band),  the percentage of objects not
detected for the plane and bulge are comparable.  We interpret this as that,
on average, the bulge images show better contrast than the plane ones.

In general, extended PNe at H$\alpha$ are related to unresolved or small
extended (PSF$_{\rm PN} \ge$ PSF$_{\rm field}$) sources in the NIR images.  In
these cases, it is probable that we are measuring the central star of the PN,
or emission from its immediate environment. The coordinates are shown in
Table~\ref{tablita}.

Our catalogue (Table~\ref{tablita}) includes 353 PNe with $ZYJHK_{\rm s}$
photometrical data.  From these, 209 are new NIR detections. The remaining
ones were previously detected and measured by 2MASS, however the VVV
photometry improves on these earlier determinations.

%%%%%%%%%%%%%%%%%%%%%%%%%%%%%%%%%%%%%%%%%%%%%%%%%%%%%%%%%%%%%%%%%%%%%%%%%%%%%%%%
%%%%%%%%%%%%%%%%%%%%%%%%%%%%%%%%%%%%%%%%%%%%%%%%%%%%%%%%%%%%%%%%%%%%%%%%%%%%%%%%
\subsection{Comparison between VVV and 2MASS}
%%%%%%%%%%%%%%%%%%%%%%%%%%%%%%%%%%%%%%%%%%%%%%%%%%%%%%%%%%%%%%%%%%%%%%%%%%%%%%%%
%%%%%%%%%%%%%%%%%%%%%%%%%%%%%%%%%%%%%%%%%%%%%%%%%%%%%%%%%%%%%%%%%%%%%%%%%%%%%%%%

To compare the VVV magnitudes with those of 2MASS 
(The 2MASS Point Source catalogue) ones,
we selected objects with
\textit{J(2MASS)$<$16.0} and \textit{J(VVV)$>$11.8},
\textit{H(2MASS)$<$15.5} and \textit{H(VVV)$>$11.0},
\textit{K$_{\rm  s}$(2MASS)$<$14.0} and \textit{K$_{\rm  s}$(VVV)$>$11.0},
according to the upper and lower detection limit, respectively.
The brightest objects in VVV were not taken into account
since they are near the saturation limit.
We also did not consider the weakest PNe in the 
2MASS survey because of the uncertainties on the photometry.

\begin{figure}
   \centering
   \includegraphics[width=8cm]{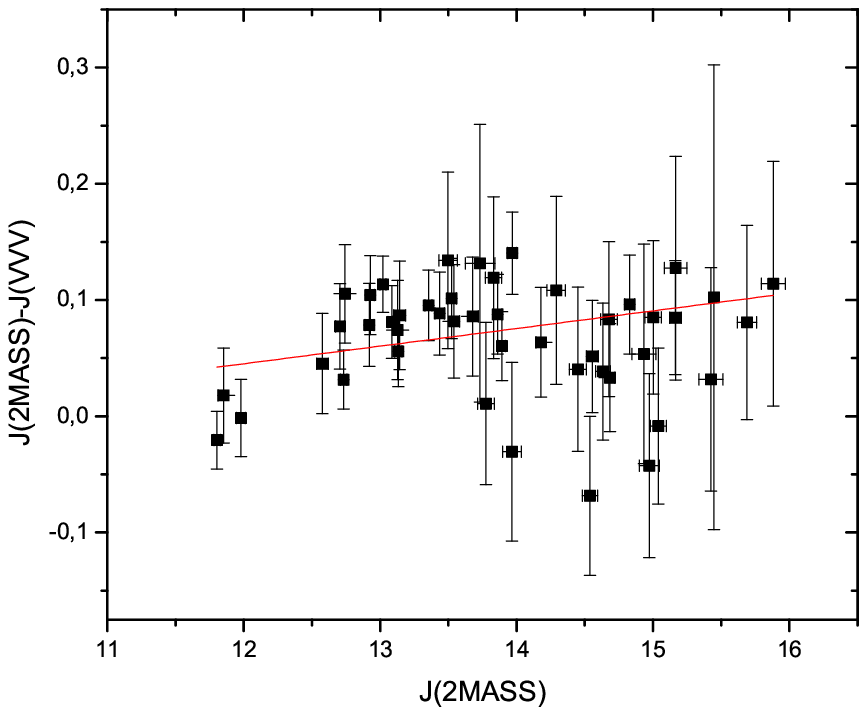}
   \includegraphics[width=8cm]{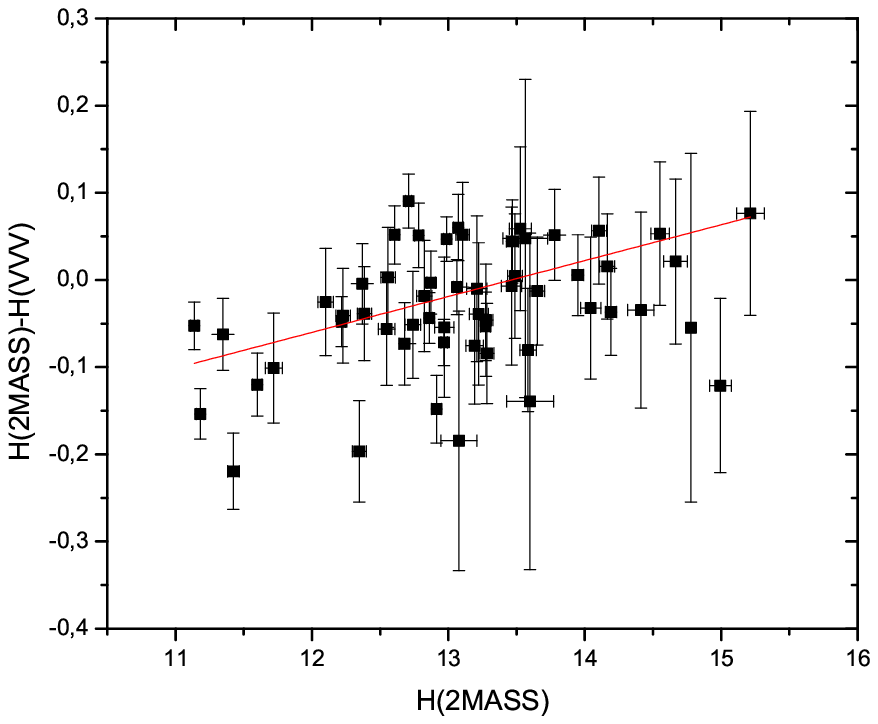}
   \includegraphics[width=7.5cm]{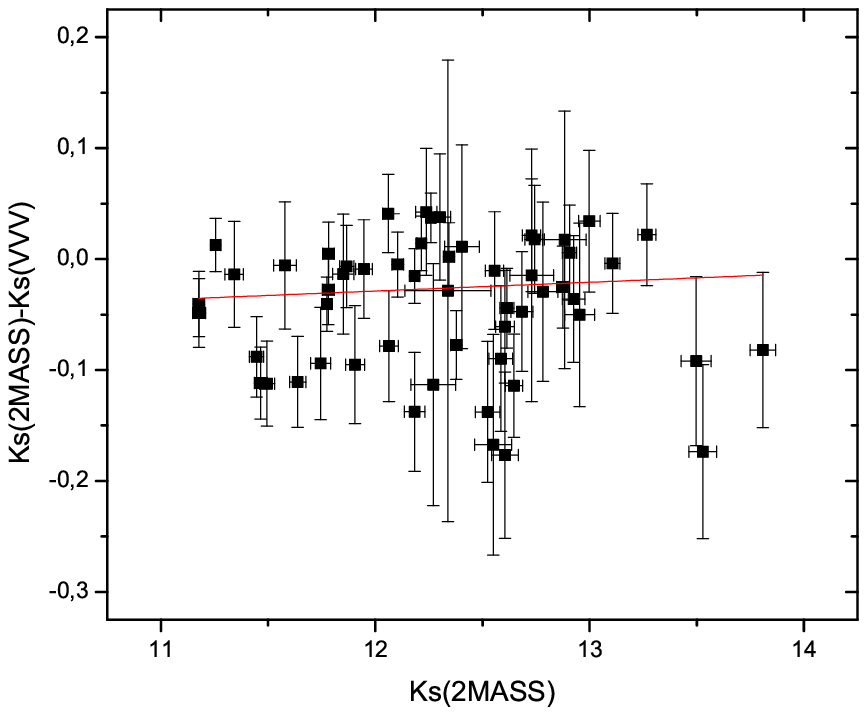}
      \caption{Comparison between 2MASS and VVV PNe magnitudes.}
         \label{j2-jv}
   \end{figure}

The 2MASS Point Source Catalogue was constructed using aperture photometry with
4$\arcsec$ radius \citep{2006AJ....131.1163S}, while we used aperture
photometry with a 1.41$\arcsec$ radius.  The difference between both
magnitudes is about 0.07 mag at J and lower for the other bands (see
Table~\ref{vvv2mass} and Fig.~\ref{j2-jv}), therefore we did not apply any
additional corrections to the VVV data.
This is because the VISTA and 2MASS photometric systems are different, 
even if the filter names are the same.
The VISTA magnitude of an object is not
expected to be exactly the same as the 2MASS magnitude 
(except for an unreddened A0V star) and vice versa.

Examining VVV images showed that the detected NIR emission of the PNe, in
general, is point-like or nearly so. The larger aperture of the 2MASS
photometry will therefore have little effect on the flux of the object, but it
can lead to increased contamination by neighbouring stars, especially in
crowded fields (very frequent for the bulge region).

%%%%%%%%%%%%%%%%%%%%%%%%%%%%%%%%%%%%%%%%%%%%%%%%%%%%%%%
%%%%%%%%%%%%%%%%%%%%%%%%%%%%%%%%%%%%%%%%%%%%%%%%%%%%%%%
\subsection{The analysis of the VVV photometry}
%%%%%%%%%%%%%%%%%%%%%%%%%%%%%%%%%%%%%%%%%%%%%%%%%%%%%%%
%%%%%%%%%%%%%%%%%%%%%%%%%%%%%%%%%%%%%%%%%%%%%%%%%%%%%%%

\subsubsection{The (\textit{J$-$H}) vs. (\textit{H$-$K$_{\rm s}$}) diagram}

\begin{figure}
   \centering
   \includegraphics[width=9cm]{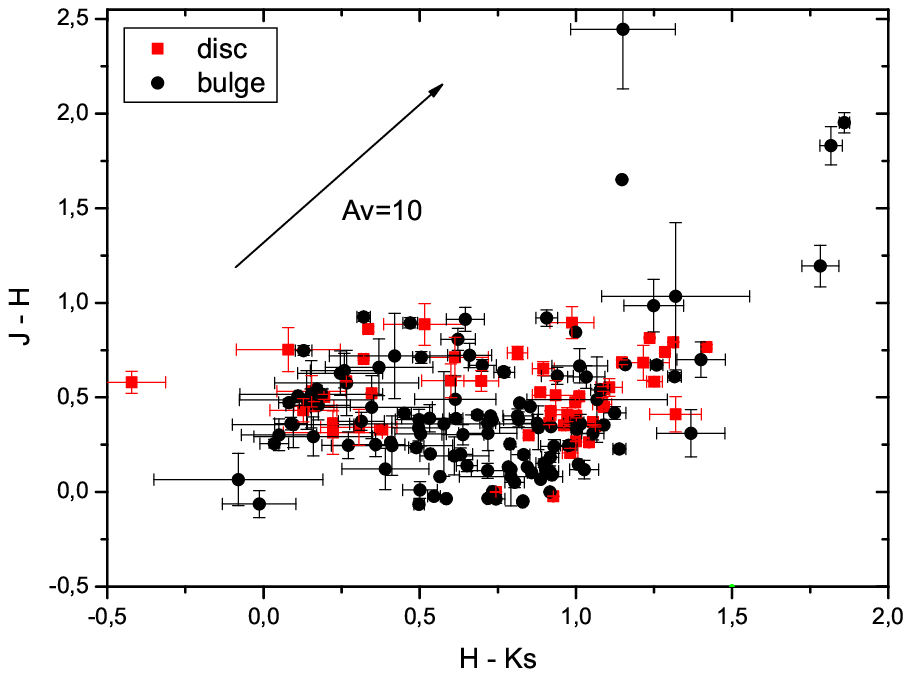}
   \includegraphics[width=9cm]{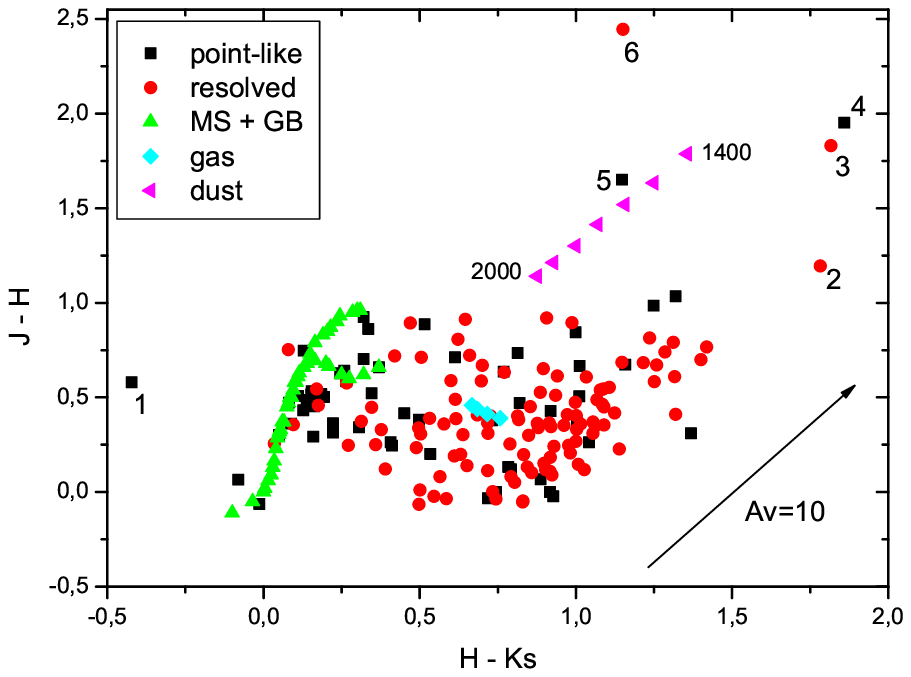}
      \caption{The upper panel shows the distribution of PNe according to
        whether they belong to the plane or bulge. The lower panel shows the
        distribution of colour indices (\textit{$H-K_{\rm s}$}) and (\textit{$J-H$}) for
        the PNe analysed as 'galaxy-like' (resolved) by the pipeline, plotted
        as red circles, point-like sources, plotted as black squares; error
        bars have been left our for the sake for clarity.  The magenta
        triangles indicate dust emission loci for an emissivity exponent
        $\gamma$=0 and temperature T$_{d}\leqslant$2000 K (in steps of 100 K,
        \citealp{1994A&AS..104..169P}).  Intrinsic stellar indices for MS
        spectral types \citep{1988PASP..100.1134B} and the O-type from
        \cite{2005A&A...436.1049M} are shown by green triangles. Finally, the
        free-free and free-bound emission from hydrogen ions are shown for the
        range of 8400 to 17000 K, by the cyan diamond symbols.  }
         \label{dedo}
   \end{figure}

All known PNe with photometric data in the VVV are placed in the
\textit{J$-$H} vs. \textit{H$-$K$_{\rm s}$}
diagram (Fig.~\ref{dedo}).  More than 50\% are objects
with original NIR data.  The magnitudes of the sources are listed in
Table~\ref{tablita}.  The magnitudes are not dereddened; the reddening vector,
obtained from \cite{2012A&A...537A.107S} is indicated by the arrow (its length
corresponds to 10 mag extinction in V).

PNe belonging to the bulge and disc are plotted separately in the upper
panel. The lower panel distinguishes point-like from extended objects, as
classified by the pipeline.  The interpretation of these diagrams was already
discussed in \cite{2008A&A...480..409C}, and our results are consistent
(\citealt{2005MNRAS.357..732R}, \citealt{1997A&AS..126..479G},
\citealt{2009MNRAS.394.1875P}). The expected colours for unreddened main
sequence and giant branch stars are shown in the lower diagram, extending from
Vega at (0, 0) (in the vicinity of the Rayleigh-Jeans point) to 
(\textit{H$-$K$_{\rm s}$}, \textit{J$-$H}) =(0.5, 1.0). 
S-type symbiotic stars (and other emission line stars) are located
around (0.5, 1.0), and for D-type symbiotics and T Tau stars a hot dust
continuum is added to those to give a broad sequence towards (2, 2).
Fig.~\ref{dedo} shows that the majority of the VVV PNe are located well away
from this, and relatively few are found among the stellar locus.

Compared to the sample of \cite{2008A&A...480..409C}, there is little evidence
for significant further reddening. This is not unexpected as the PNe in our
sample had previously been discovered in optical surveys.

To investigate if some difference between PNe populations from disc and bulge
is present, we distinguished them in the upper panel of Fig.~\ref{dedo}. There
is an indication for a separation, with the bulge PNe (black dots) slightly
bluer both in \textit{J$-$H} and \textit{H$-$K$_{\rm s}$} for objects around the point 
(\textit{$H-K_{\rm s}$}, \textit{$J-H$}) = (0.7, 0.5) where many of the PNe are found. The
separation is much less evident among the objects with stellar colours
(presumed field stars). This suggests that the separation is not due to
systematic differences in extinction.

As mentioned in Section~\ref{dbase}, the VVV pipeline distinguish between
stars (point sources) and 'galaxies' (i.e. not point sources).  The PNe,
according to the flag point-like/not-point-like derived from the VVV pipeline,
are plotted in the lower panel of Fig.~\ref{dedo}.  In general, the analysed
sources are interpreted by the pipeline as resolved (66\% of the sample shown
in Fig.~\ref{dedo}). This could be an interesting aid to find new PNe.  The
distribution of sources within the \textit{$J-H$} vs. \textit{$H-K_{\rm s}$} plane
(Fig.~\ref{dedo}, lower panel) shows two concentrations of sources, but the
distinction between the stellar and nebular sources only marginally reflects
the unresolved/resolved flag. The relation between this flag and the nature of
the emission is not obvious.

Interestingly, the stellar locus is sparsely populated  compared to other
works (for instance \citealp{2009MNRAS.394.1875P}).  This fact perhaps shows
that VVV, with better resolution, yields improved rejection of unrelated field
stars.  It is noteworthy that there is a distinct concentration near the locus
of unreddened K2 stars.  This may indicate these are indeed field stars.

The lower panel of Fig.~\ref{dedo} also shows the loci of free-free/bound-free
emission from ionised hydrogen gas. These are calculated for a temperature
range from 8400 to 17000 K, at approximately (\textit{$H-K_{\rm s}$}, \textit{$J-H$})=(0.7, 0.4). 
 The actual measurements deviate from this by about 0.5 mag, especially
in \textit{$H-K_{\rm s}$}. This is mainly due to contributions from emission
lines. Interestingly, the bulge PNe tend to be located closer to the bf/ff
locus, perhaps reflecting a different emission-line contribution.

\begin{table}
\caption{VVV and 2MASS magnitude comparison (linear fit).}             
\label{vvv2mass}      
\centering                          
\begin{tabular}{c c c c c}        
\hline\hline                 
  & N & average & RMS & slope \\   
\hline                        
   \textit{J(2MASS)$-$J(VVV)}  & 46 &   +0.066  & 0.049   & 0.015 $\pm$ 0.007   \\      
   \textit{H(2MASS)$-$H(VVV)}  & 56 & $-$0.032  & 0.071   & 0.041 $\pm$ 0.010 \\
   \textit{K$_{\rm s}$(2MASS)$-$K$_{\rm s}$(VVV)} & 55 & $-$0.042  & 0.059   & 0.008 $\pm$ 0.010 \\
\hline                                  
\end{tabular}
\end{table}

We finally compare the observed distribution of PNe with the model of NGC
6720. This gives colours of (\textit{$H-K_{\rm s}$}, \textit{$J-H$}) = (0.62, 0.0), 
towards the bottom range of the observed colours. For a
similar model for NGC\,7027 \citep{1996A&A...315L.253B, 2008ApJ...681.1296Z}
we derive unreddened colours of (\textit{$H-K_{\rm s}$}, \textit{$J-H$})=(0.21, 0.41)
also within the general PNe region but close to the blue edge of the PNe
distribution. Allowing for a range of extinction values in the observed
sample, these two models can cover much of the observed range.

%%%%%%%%%%%%%%%%%%%%%%%%%%%%%%%%%%%%%%%%%%%%%%%%%%%%%%%
%%%%%%%%%%%%%%%%%%%%%%%%%%%%%%%%%%%%%%%%%%%%%%%%%%%%%%%
\subsubsection{The (\textit{Y$-$J}) vs. (\textit{Z$-$Y}) diagram }
%%%%%%%%%%%%%%%%%%%%%%%%%%%%%%%%%%%%%%%%%%%%%%%%%%%%%%%
%%%%%%%%%%%%%%%%%%%%%%%%%%%%%%%%%%%%%%%%%%%%%%%%%%%%%%%

\begin{figure}
   \centering
   \includegraphics[width=9cm]{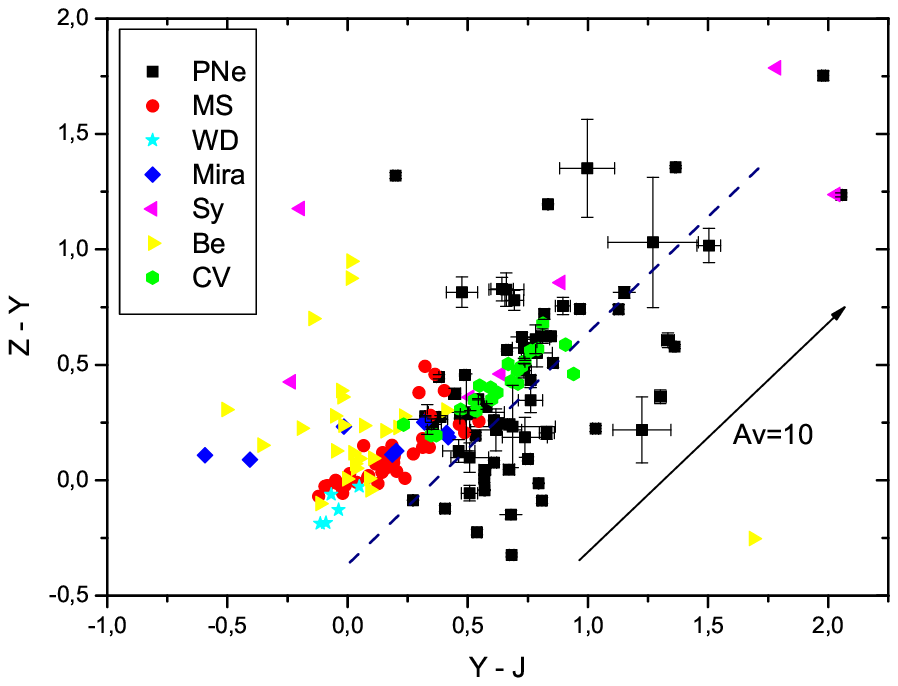}
   \includegraphics[width=9cm]{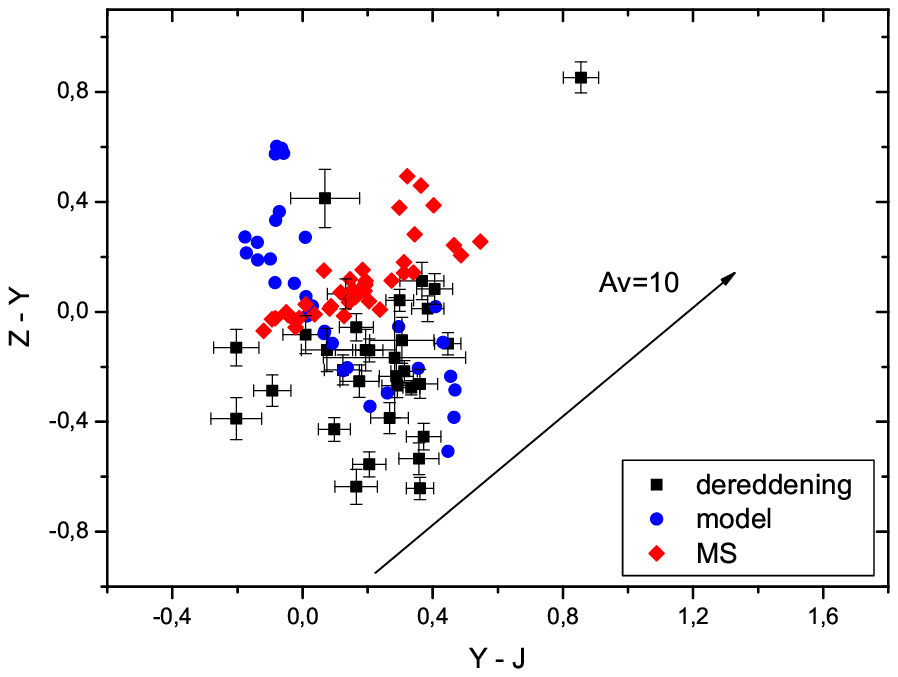}
      \caption{Distribution of the colour indices (\textit{$ Y-J$}) and (\textit{$Z-Y$}).
        The upper panel show all detected PNe without applying any extinction
        correction. Other symbols show other emission-line stars, main
        sequence stars \citep{2006MNRAS.367..454H} and white dwarfs (taken
        from the standard database of MKO) as indicated.  The bottom panel
        shows dereddened colours for the PNe, using the reddening measured
        from the hydrogen lines.  The distribution of the colour indices 
        (\textit{$Y-J$}) and (\textit{$\ Z-Y$}) for a model grid of planetary nebulae is shown
        in blue.}
   \label{zyj}
   \end{figure}

Figure~\ref{zyj} shows the distribution of PNe in the newly measured 
(\textit{$Y-J$}) vs. (\textit{$Z-Y$}) diagram. We analysed the reddened and dereddened colours
(upper and lower panel respectively).  In addition, the colours of other
emission objects are shown, to be compared with those of PNe.  The reddening
vector, obtained from \cite{2012A&A...537A.107S}, is indicated by an arrow
(its length corresponds to 10 mag extinction in V).

We have extracted from the VVV database the $Z$, $Y$ and $J$ magnitudes of
symbiotic stars \citep{2000A&AS..146..407B}, cataclysmic variables
\citep{2001PASP..113..764D}, Be stars \citep{2005NewA...10..325Z}, and Mira
variables \citep{2002A&A...384..925K}.  Although the samples of these objects
are not very large, they are useful to estimate the possible overlap with PNe.
The distribution of reddened PNe (Fig.~\ref{zyj}, upper panel) occupies a wide
region in the (\textit{$Y-J$}) vs. (\textit{$Z-Y$}) diagram, but it is possible to
isolate an area to the right of the main sequence (MS) and parallel to the
reddening vector (limited by the dashed line), which is not shared by the
other objects.  Although only 41\%\ of PNe are placed to the right of the
dashed line, the lack of confusion in this region suggests that this diagram,
combined with the standard $JHK_{\rm s}$, will be a useful tool for the
identification of new PNe.

The $ZYJ$ diagnostic diagram, using dereddened colours, is shown in
Figure~\ref{zyj} (bottom panel), together with the location of MS stars.  The
colours are corrected for the interstellar reddening using, in general, the
extinction coefficient $c$ of \cite{1992A&AS...95..337T}, and by assuming that
\textit{$c/E_{\rm B-V}$} = 1.46
\citep{1984ASSL..107.....P} and taking the standard 
\textit{$A_{}$/$E_{B-V}$} = 3.1.  The infrared extinction $A_{Z} = 0.499A_{V}$, $A_{Y} =
  0.390A_{\rm V}$ and 
$A{\rm _J} = 0.280A_{\rm V}$ were taken from
  \cite{2012A&A...537A.107S}.  We estimate an uncertainty of 10\% on the
  extinction coefficients $c$ (see Fig.~\ref{zyj}).

There is some overlap in colours between MS stars and unreddened PNe, and this
overlap is increased by reddening (top panel). This diagram does not allow us
to differentiate PN from MS stars with high efficiency. However, it is
evident that the PNe tend to be grouped in a bounded region, and there is a
well-defined region containing about half the PNe which is not affected by
confusion with the depicted types of objects.

The $Y$ and $Z$ bands have strong contributions from emission lines
and have less continuum as the bands are relatively narrow (as compared to
$JHK_{\rm s}$). The $J$-band contains a strong Pa$\beta$ line but also
more continuum, as the band is wider. The \textit{$Z-Y$} colour can become negative
because the $ Z$-band contains a strong [S{\sc iii}] line.

%%%%%%%%%%%%%%%%%%%%%%%%%%%%%%%%%%%%%%%%%%%%%%%%%%%%%%%
%%%%%%%%%%%%%%%%%%%%%%%%%%%%%%%%%%%%%%%%%%%%%%%%%%%%%%%
\subsubsection{Notes on individual objects} \label{secti}
%%%%%%%%%%%%%%%%%%%%%%%%%%%%%%%%%%%%%%%%%%%%%%%%%%%%%%%
%%%%%%%%%%%%%%%%%%%%%%%%%%%%%%%%%%%%%%%%%%%%%%%%%%%%%%%

\begin{figure*}[!ht]
\centering
\includegraphics[width=5cm]{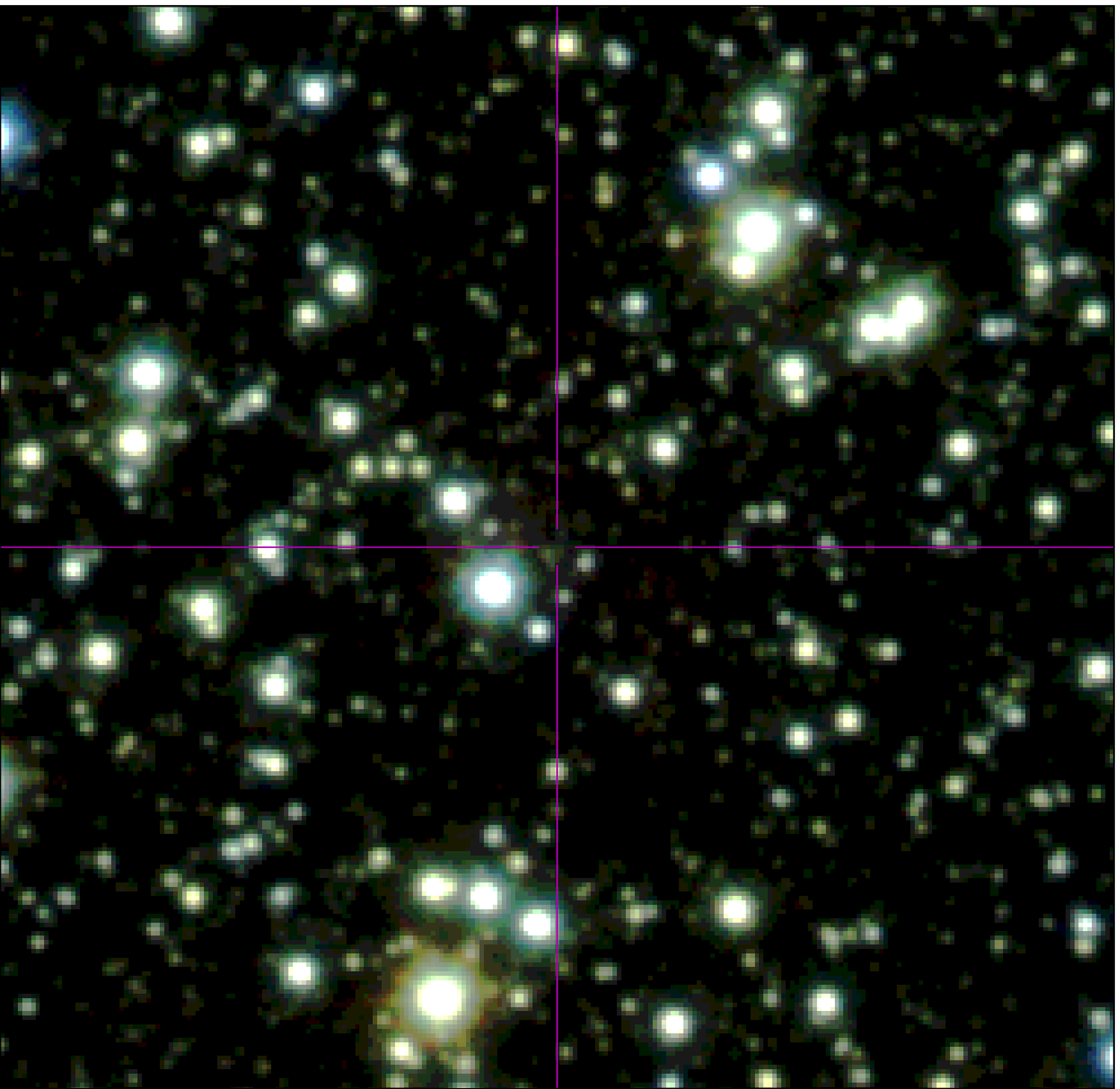}
\includegraphics[width=5cm]{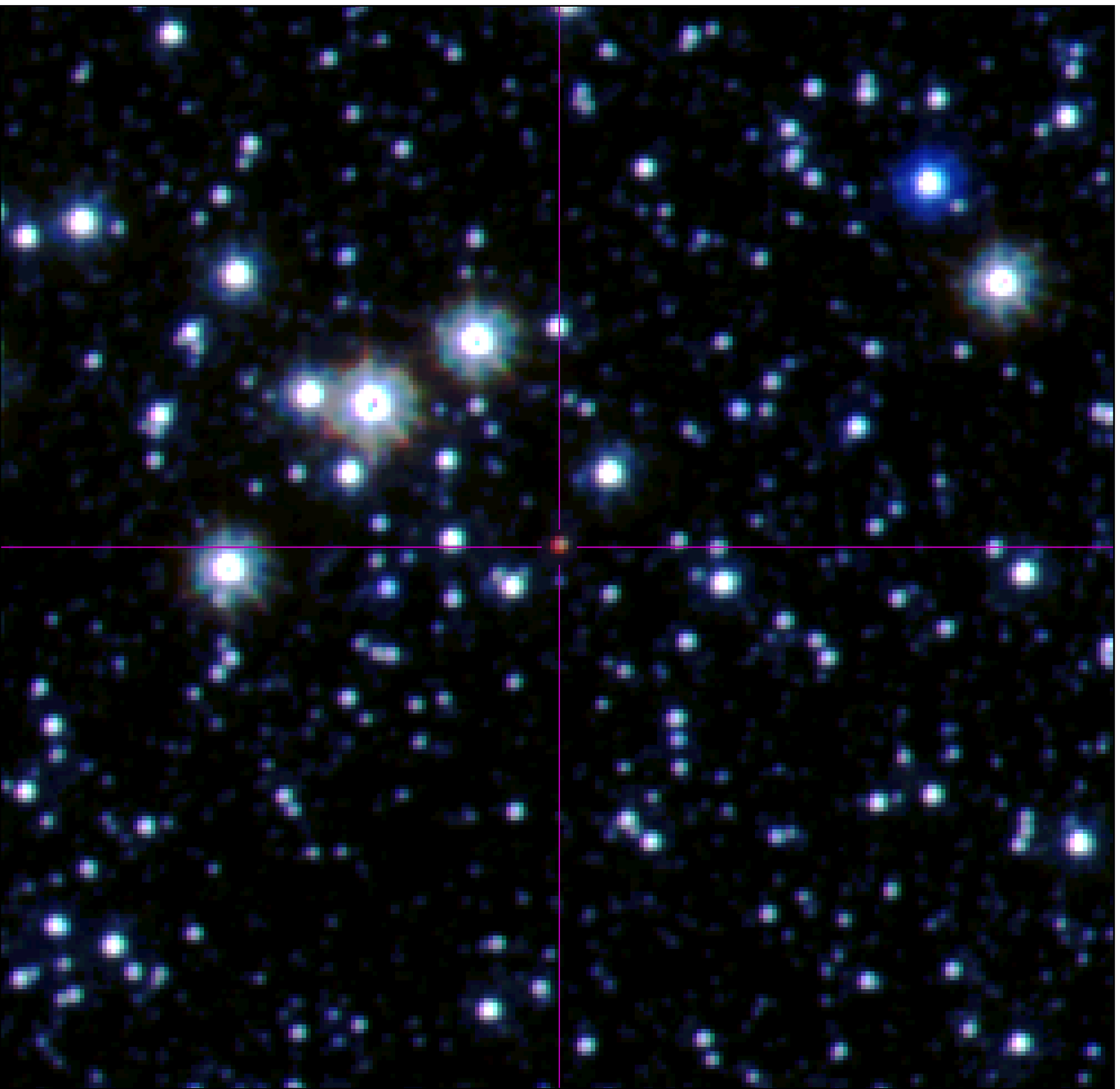}
\includegraphics[width=5cm]{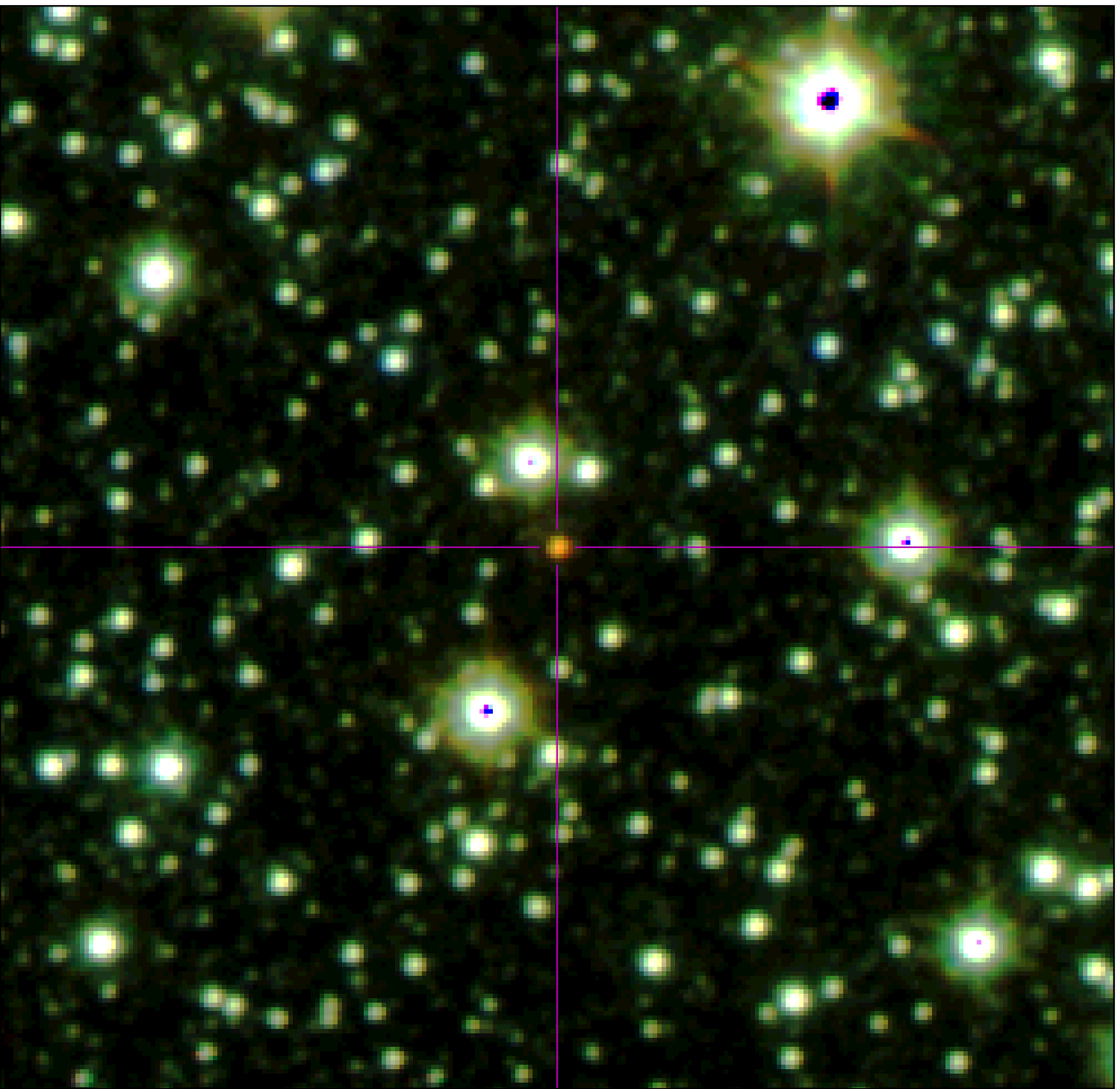}
\includegraphics[width=5cm]{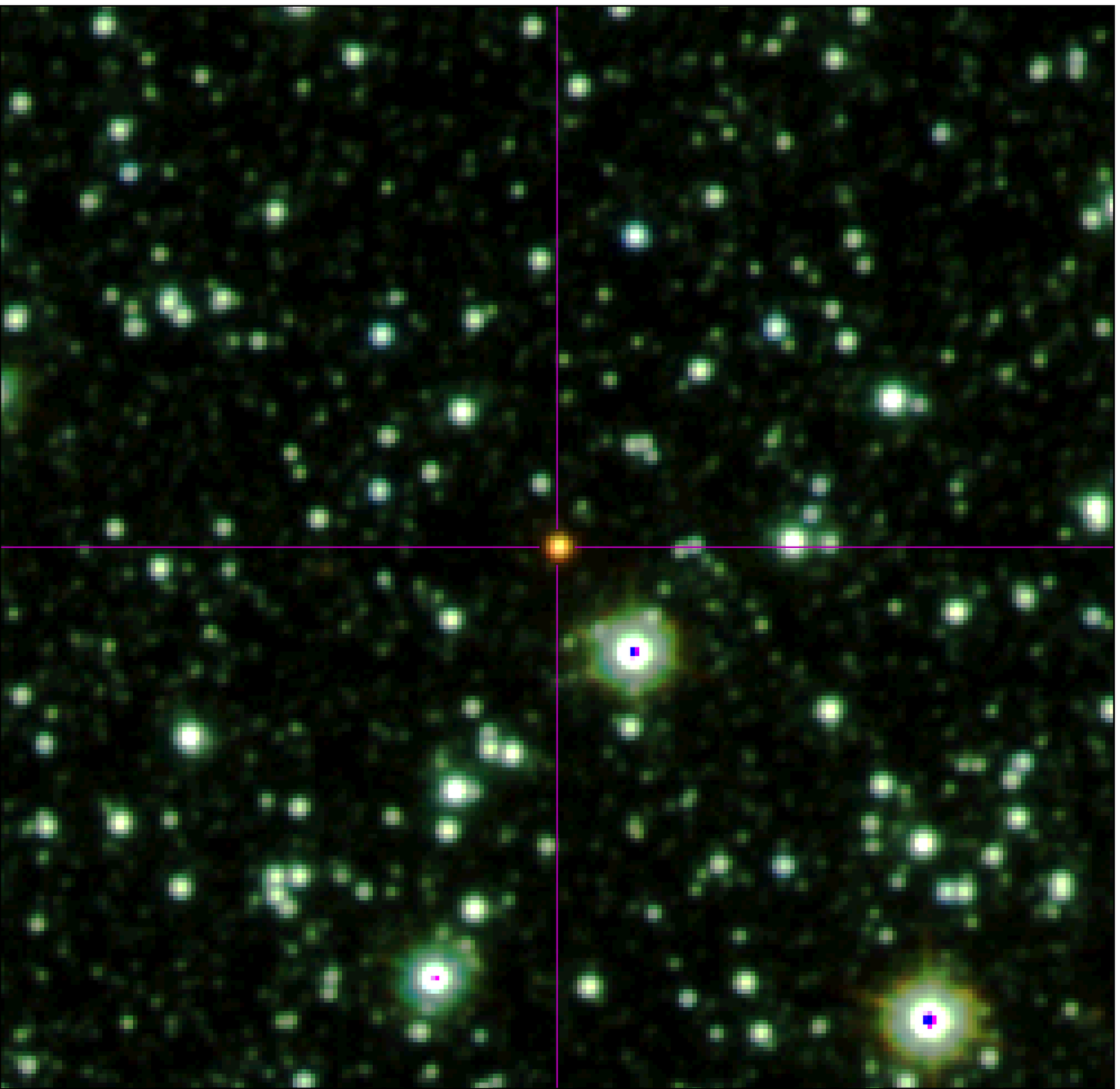}
\includegraphics[width=5cm]{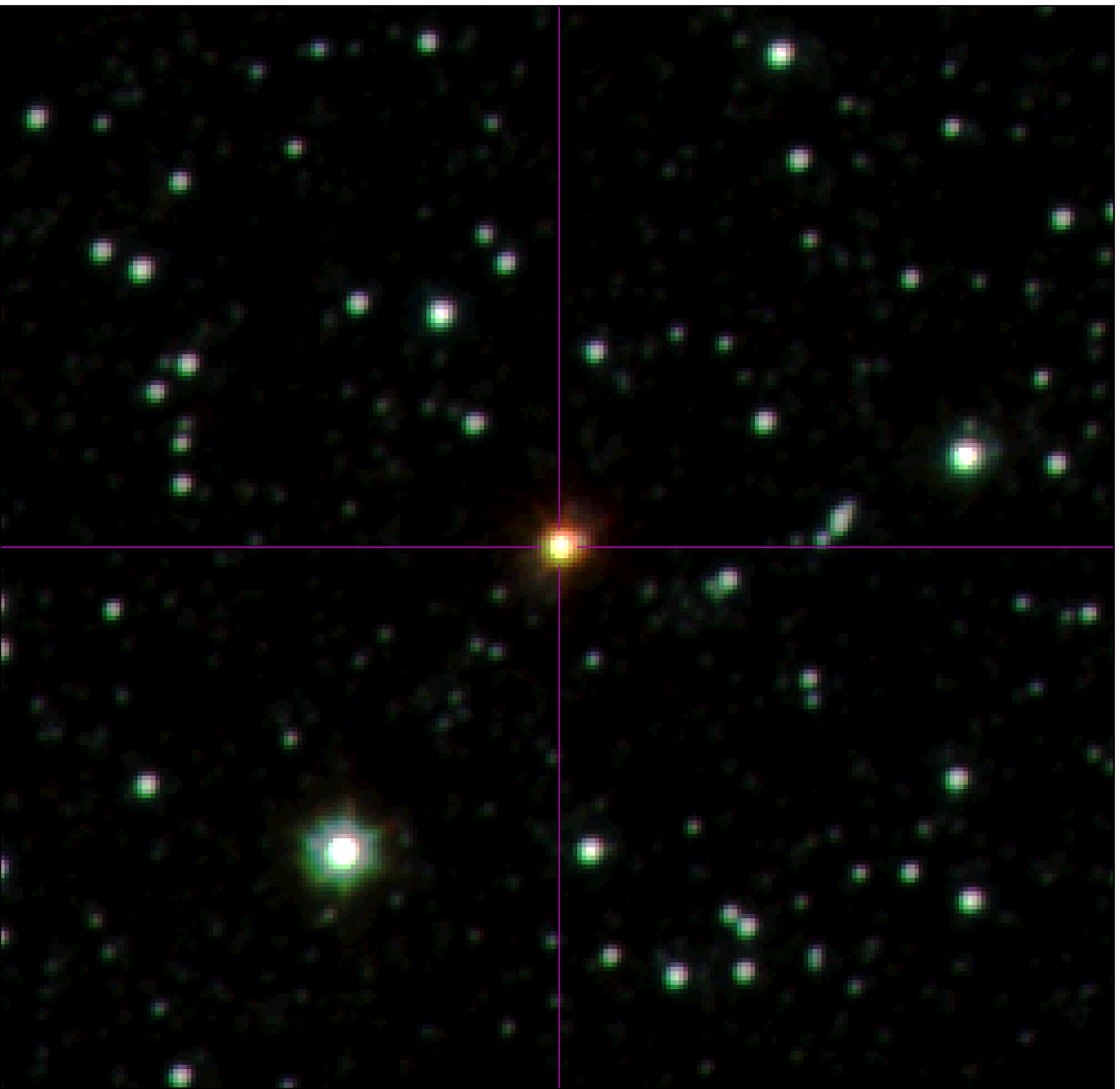}
\includegraphics[width=5cm]{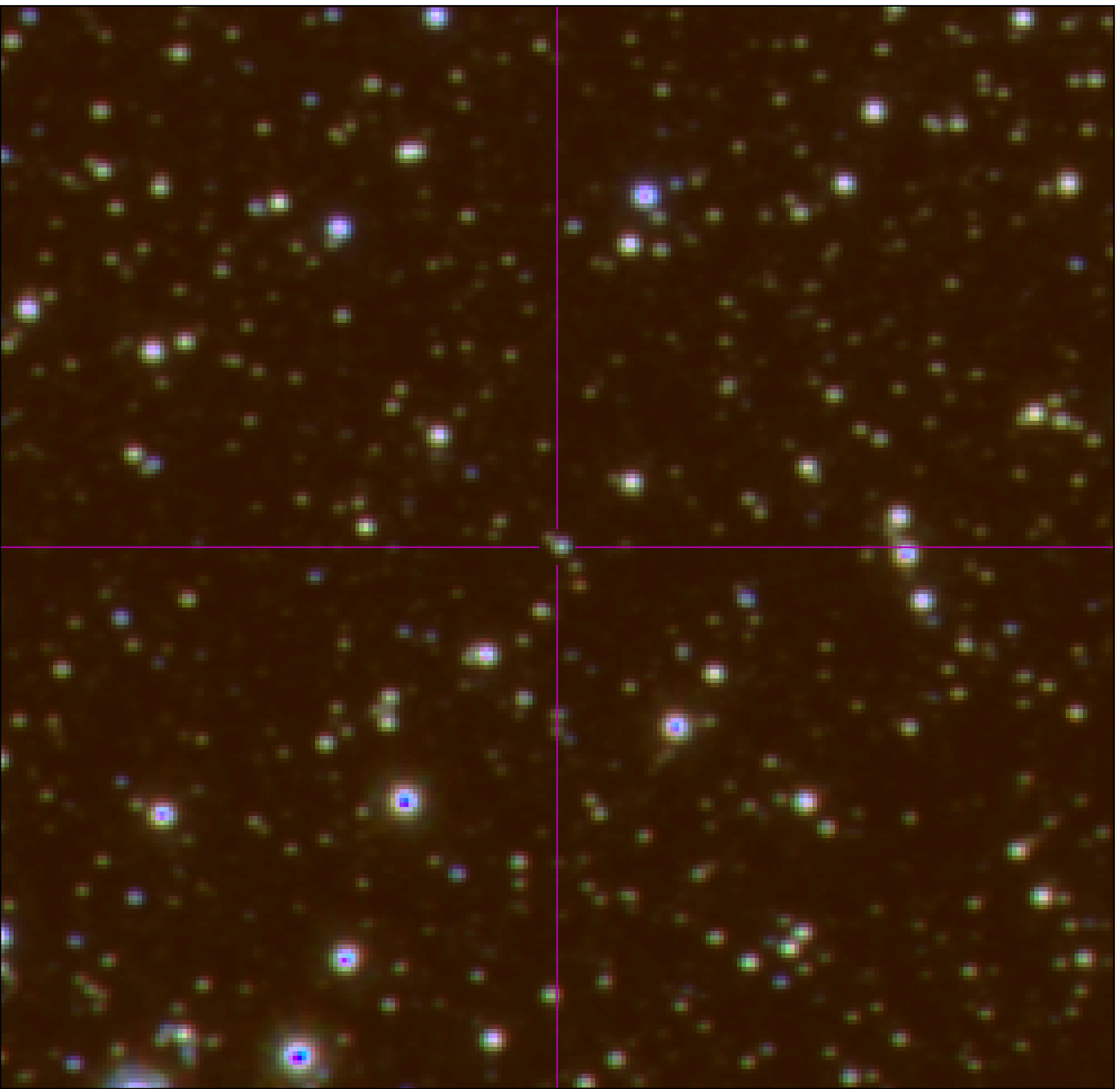}
  \caption{$JHK_{\rm s}$ colour composite images showing the six 
stars (red cross) described at section~\ref{secti}, range from top left to
bottom right. The size of images are 1.4'.}
   \label{imagenes}
\end{figure*}

A number of objects show unexpected colours. These are labelled by number in
Fig.~\ref{dedo}, lower panel. The label number is indicated below, after the
name of the PN (see Fig.~\ref{imagenes}).

\paragraph{ {PN G$347.4+01.8$ }.--- (1)}
This shows a very blue (\textit{$H-K_{\rm  s}$}) colour.
The H$\alpha$ image shows a bipolar morphology \citep{2006MNRAS.373...79P}, 
we identify a VVV NIR source at the geometric centre of the belt, 
however this could be a spurious star.

\paragraph{{PN G$001.9+02.3$}.--- (2)}
The colour is indicative of a highly reddened PN.
However, we could not reject a very young PN 
\citep{2009A&A...501.1207R, 2006IAUS..234..449L}.

\paragraph{{PN G$355.2-02.0$}.--- (3)}
Similar to the previous object, the colour indicates high reddening. 
The object has a high excitation class 
\citep{2006MNRAS.373...79P}.
However, in agreement with \citet[Fig.~1]{2009A&A...502..113V} 
this could be a D-type symbiotic star.

\paragraph{{PN G$355.0-03.3$}.--- (4)}
Suspected to be a symbiotic star \citep{2009A&A...496..813M}.

\paragraph{{PN G$356.9-05.8$}.--- (5)}
Its colours can be
explained as dominated by hot dust.
It is suspected to be a symbiotic star \citep{2009A&A...496..813M}.

\paragraph{{PN G$000.1-01.0$}.--- (6)}
This object has unusual colours. It is a poorly studied object, 
and perhaps it is not a PN.

%%%%%%%%%%%%%%%%%%%%%%%%%%%%%%%%%%%%%%%%%%%%%%%%%%%%%%%
%%%%%%%%%%%%%%%%%%%%%%%%%%%%%%%%%%%%%%%%%%%%%%%%%%%%%%%
\subsection{Model distribution}
%%%%%%%%%%%%%%%%%%%%%%%%%%%%%%%%%%%%%%%%%%%%%%%%%%%%%%%
%%%%%%%%%%%%%%%%%%%%%%%%%%%%%%%%%%%%%%%%%%%%%%%%%%%%%%%

We can again compare the colours of the NGC\,6720 model with the observed
distribution. The synthetic colour of (\textit{$Y-J$}, \textit{$Z-Y$})=(0.12, 0.21). The
NGC 7027 model has very similar (unreddened) colours of (\textit{$Y-J$}, 
\textit{$Z-Y$})=(0.10, 0.24).  These fall at the edge of the observed distribution
(upper panel of Fig.~\ref{zyj}).
NGC\,7027 is a young, high density PN, whilst NGC\,6720 is evolved and has low
density.  It is therefore not immediately obvious what causes any 
differences with the observed distribution.

To explore this further, we computed a Cloudy model grid, and for each model
in the grid calculated the \textit{$Y-J$} and \textit{$Z-Y$} colours. The grid is defined
by varying the effective temperature of the star and the hydrogen density of
the nebula. The grid runs from \textit{$\log T_{\rm eff}= 4.7$} to 5.4 in steps of 0.1
dex (temperatures from 50 kK to 250 kK), and from \textit{$\log n_{\rm H}=2$} to 5 in
steps of 1 dex. The model nebula has exactly $N_{\rm H}= 10^{57}$ atoms
corresponding to a mass of approximately 1.2\,M$_\odot$, and all models are
calculated up to the outer radius with inclusion of molecular hydrogen in a
PDR (Photon Dominated Regions) region, if present.

The lower panel of Fig.~\ref{zyj} shows the distribution of points in the
(\textit{$Y-J$}) and (\textit{$Z-Y$}) plane.  The models follow the lower edge of the
observed PNe distribution in the upper panel of Fig.~\ref{zyj}. Together with
a range of extinction in the VVV sample, the observed distribution can be well
explained.  There is some discrepancy in the lower panel, in that some
dereddened data fall below the model grid. This may reflect limitations in the
model grid, such as a constant nebular mass and constant abundances
(especially for sulphur). It is also possible that in some
cases the extinction coefficient $c$ of \cite{1992A&AS...95..337T} is
over-estimated, or that the scaling from H$\alpha$/H$\beta$ to optical and
near-infrared broad-band colours is also affected by the shape of the reddening
law. \citet{2004MNRAS.353..796R} have shown indications that the common
assumption of $R=3.1$ used here may not always be appropriate for known PNe.

\begin{figure}
   \centering
   \includegraphics[width=9cm]{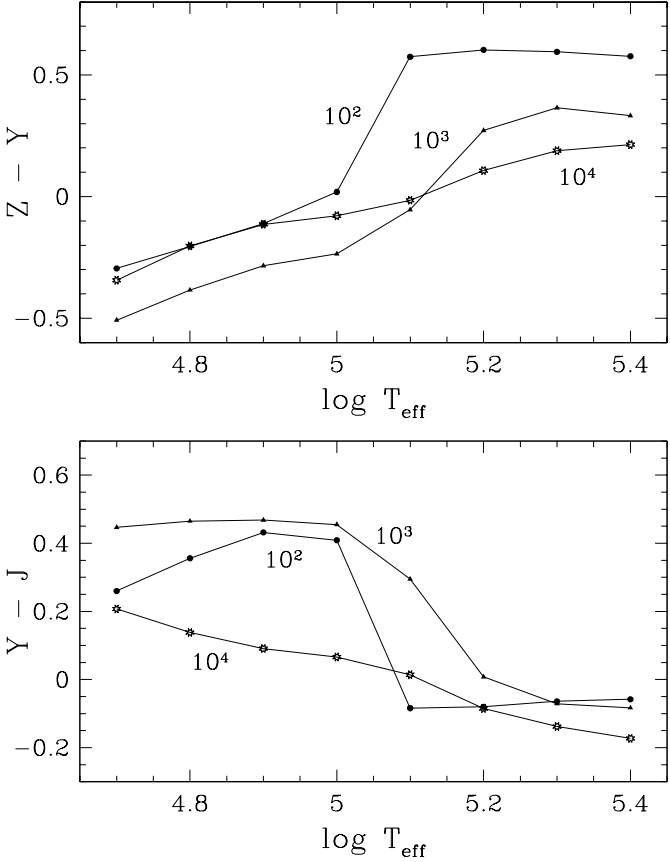}
      \caption{Distribution of the colour indices (\textit{$Y-J$}) and (\textit{$Z-Y$}) as
        function of stellar temperature, for the Cloudy model grid described
        in the paper. Each curve is for a fixed density, indicated by the
        adjacent label.  }
   \label{temp}
   \end{figure}

The model grid extends to a region above the stellar locus in the (\textit{$Y-J$},
\textit{$Z-Y$}) diagram. There are almost no observed PNe in this region where many
of the models fall. These models are those with hottest ionizing stars. At all
densities, the \textit{$Z-Y$} colour increases with increasing stellar
temperature, as shown in Fig. \ref{temp}. The largest change occurs between
temperatures of $10^5$ and $1.25 \times 10^5\,$K. The \textit{$Y-J$} curve is
largely the mirror image of the \textit{$Z-Y$} curve. This indicates that the
change is caused by a high excitation line in the $\rm Y$ band.  The He\,{\sc
  i} 1.083$\mu$m lines decreases between $T_{\rm eff}=10^5$ en $10^{5.2}$\,K, and
the He\,{\sc ii} 1.012$\mu$m becomes much stronger. Both lines are in the Y band,
but the latter dominates because of the much higher transmission efficiency at
its wavelength (Table \ref{n6720_lines}). The effect is strongest for the
lower density models as these are density bounded, and lack any lower
ionisation regions. Note that the model in Table \ref{n6720_lines} is
ionisation bounded.

The four model points at (\textit{$Y-J$}, \textit{$Z-Y$})$ = (-0.08, 0.55)$ are low
density models with $n_{\rm H} =10^2\,\rm cm^{-3}$. These are fully ionised
and have very large predicted diameters ($22\,$arcsec) at the distance of the
bulge. The three models just below this have $n_{\rm H} =10^3,\rm cm^{-3}$,
are also fully ionised with predicted diameters of $10\,$arcsec.  The nebulae
which are missing from the observed distribution are those with very hot stars
and large to very large nebulae.  
Such nebulae may have been removed from the
sample (as described above), may not have been incuded in the VVV catalogues,
or their colours have been affected by the use of an aperture much smaller than
the size of the nebula.

%%%%%%%%%%%%%%%%%%%%%%%%%%%%%%%%%%%%%%%%%%%%%%%%%%%%%%%
%%%%%%%%%%%%%%%%%%%%%%%%%%%%%%%%%%%%%%%%%%%%%%%%%%%%%%%
\section{Conclusions}
%%%%%%%%%%%%%%%%%%%%%%%%%%%%%%%%%%%%%%%%%%%%%%%%%%%%%%%
%%%%%%%%%%%%%%%%%%%%%%%%%%%%%%%%%%%%%%%%%%%%%%%%%%%%%%%

The VVV survey is shown to be sensitive to planetary nebulae. Of the 579 known
PNe (excluding 75 highly extended objects) in the covered region, 353 have
been detected in at least one filter. The majority of these (209) are new
near-infrared detections. The NIR images, to the naked eye, suggest compact,
point-like sources; however the VVV pipeline classifies a high percentage of
PNe as extended sources. Aperture photometry was obtained from the pipeline
catalogues. In the J, H, K$_s$ bands, the photometry agrees well with 2MASS data
where available, in most cases within 0.1 mag,
even though 2MASS and VISTA are each on their own photometric system.

We use a photo-ionisation model of NGC 6720 (the Ring nebula) to calculate the
flux contributions from emission lines and from the continuum, for each of the
five filters. The line contribution decreases with increasing wavelength, from
55\%\ in $\rm Z$ and $\rm Y$, to 22\% in $\rm K_s$. The precise contribution
depends on the excitation of the nebula, especially at $\rm Z$ and $\rm Y$
where the dominant lines are from sulphur and helium, respectively. The $\rm
J$, $\rm H$, $\rm K_s$ bands are dominated by hydrogen, where it should be
noted that molecular hydrogen can also contribute at $\rm K_s$. Electron
temperature and elemental abundances may also affect the relative
contributions.

We explore two diagnostic diagrams: \textit{$H-K_s$} vs \textit{$J-H$}, and \textit{$Y-J$} vs
\textit{$Z-Y$}. In the former, the PNe are well separated from the stellar locus
and from emission line stars. The PNe are located around the colours of the
continuum from ionised hydrogen plasmas. Bulge PNe tend to be located closer
to these colours than do PNe of the Galactic disc. The reason for this is not
clear, but metallicity may play a role. The \textit{$Y-J$} vs \textit{$Z-Y$} diagram
shows a broad spread in colours, and PNe have more overlap with the stellar
locus. About half fall in a region of the diagram where there is little
confusion, so that this diagram can still be a useful tool in finding and
classifying PNe.

A model grid was calculated using the Cloudy photo-ionisation code, covering a
wide range of stellar temperatures and electron densities, and assuming
standard PN abudances. 
The model spectra were convolved with the VISTA response
functions to predict colours. The models show a range of \textit{$Y-J$} \textit{$ Z-Y$}
colours whith a sequence aproximately perpendicular to the stellar locus. The
sequence follows the lower boundary of the distribution of observed PNe; the
observed distribution can be reproduced by reddening the model nebulae. There
are no observed PNe in the region above the stellar locus, where a significant
fraction of the model nebulae are located. These are related to nebulae with
hot central stars ($T>10^5\,$K), where the He\,{\sc ii} line at
1.012$\mu$m becomes the dominant line in $Y$-band. The missing objects
correspond to very large, fully ionised nebula with hot stars which are absent
from  our selected sampe.

The reddening coefficients for many of the observed PNe have been published by
\cite{1992A&AS...95..337T}. When we use these to de-redden our sample, the
resulting distribution shows that some objects fit the models very well, but a
fraction have colours which are bluer than shown by any of the models. The
discrepancy is up to 0.4 mag in \textit{$Y-Z$} and \textit{$J-Y$}. This suggests that the
extinction in some nebulae may have been over-estimated. Alternatively, as
suggested by \citet{2004MNRAS.353..796R}, uncertainties in the $R_V$-value
(assumed to be 3.1) may have an effect.

The VVV survey is shown to be a powerful tool to study planetary nebulae in
the crowded and extincted region of the galactic plane. The large majority of
known objects ware detected, but in these regions, many PNe are expected to
have remained undiscovered. The results presented here will help locating this
missing population in the VVV database.

%%%%%%%%%%%%%%%%%%%%%%%%%%%%%%%%%%%%%%%%%%%%%%%%%%%%%%%%%%%%%%%%%%%%%%%%%%%%%%%%%%%%%%%%%%%%%%%%
%%%%%%%%%%%%%%%%%%%%%%%%%%%%%%%%%%%%%%%%%%%%%%%%%%%%%%%%%%%%%%%%%%%%%%%%%%%%%%%%%%%%%%%%%%%%%%%%
%%%%%%%%%%%%%%%%%%%%%%%%%%%%%%%%%%%%%%%%%%%%%%%%%%%%%%%%%%%%%%%%%%%%%%%%%%%%%%%%%%%%%%%%%%%%%%%%
%%%%%%%%%%%%%%%%%%%%%%%%%%%%%%%%%%%%%%%%%%%%%%%%%%%%%%%%%%%%%%%%%%%%%%%%%%%%%%%%%%%%%%%%%%%%%%%%
%%%%%%%%%%%%%%%%%%%%%%%%%%%%%%%%%%%%%%%%%%%%%%%%%%%%%%%%%%%%%%%%%%%%%%%%%%%%%%%%%%%%%%%%%%%%%%%%
%%%%%%%%%%%%%%%%%%%%%%%%%%%%%%%%%%%%%%%%%%%%%%%%%%%%%%%%%%%%%%%%%%%%%%%%%%%%%%%%%%%%%%%%%%%%%%%%

\begin{acknowledgements}

We thank the anonymous referee whose very useful remarks 
helped us to substantially improve this paper.
We acknowledge Jim Emerson for
comments that helped to improve the paper.
This research has made use of the Aladin and SIMBAD database, 
operated at CDS, Strasbourg, France.
This publication makes use of data products from the Two Micron All Sky Survey, 
a joint Project of the University of Massachusetts and IPAC/CALTECH, 
funded by NASA and NSF.
We gratefully acknowledge use of data from the ESO Public Survey programme ID
179.B-2002 taken with the VISTA telescope, data products from 
the Cambridge Astronomical Survey Unit (CASU), and from the
VISTA Science Archive at the Wide Field Astronomy Unit (WFAU). 
PvH acknowledges support from the Belgian Science Policy Office through
the ESA PRODEX program.  DM gratefully acknowledges funding from the FONDAP
Center for Astrophysics 15010003, the BASAL CATA Center for Astrophysics and
Associated Technologies PFB-06, the MILENIO Milky Way Millennium Nucleus from
the Ministry of Economy's ICM grant P07-021-F, and Proyecto FONDECYT Regular
No. 1090213 from CONICYT and the astro-Enginiering center at the Universidad
Cat\'olica (AIUC).

\end{acknowledgements}

%\begin{thebibliography}{}
\bibliographystyle{aa}
\bibliography{aa3}

\begin{thebibliography}{54}
\expandafter\ifx\csname natexlab\endcsname\relax\def\natexlab#1{#1}\fi

\bibitem[{{Acker} {et~al.}(1992){Acker}, {Marcout}, {Ochsenbein}, {Stenholm},
  {Tylenda}, \& {Schohn}}]{1992secg.book.....A}
{Acker}, A., {Marcout}, J., {Ochsenbein}, F., {et~al.} 1992, {The
  Strasbourg-ESO Catalogue of Galactic Planetary Nebulae. Parts I, II (ESO,
  Garching)}, ed. {Acker, A., Marcout, J., Ochsenbein, F., Stenholm, B.,
  Tylenda, R., \& Schohn, C.}

\bibitem[{{Allen}(1973)}]{1973MNRAS.161..145A}
{Allen}, D.~A. 1973, \mnras, 161, 145

\bibitem[{{Beaulieu} {et~al.}(1999){Beaulieu}, {Dopita}, \&
  {Freeman}}]{1999ApJ...515..610B}
{Beaulieu}, S.~F., {Dopita}, M.~A., \& {Freeman}, K.~C. 1999, \apj, 515, 610

\bibitem[{{Beintema} {et~al.}(1996){Beintema}, {van Hoof}, {Lahuis},
  {Pottasch}, {Waters}, {de Graauw}, {Boxhoorn}, {Feuchtgruber}, \&
  {Morris}}]{1996A&A...315L.253B}
{Beintema}, D.~A., {van Hoof}, P.~A.~M., {Lahuis}, F., {et~al.} 1996, \aap,
  315, L253

\bibitem[{{Belczy{\'n}ski} {et~al.}(2000){Belczy{\'n}ski}, {Miko{\l}ajewska},
  {Munari}, {Ivison}, \& {Friedjung}}]{2000A&AS..146..407B}
{Belczy{\'n}ski}, K., {Miko{\l}ajewska}, J., {Munari}, U., {Ivison}, R.~J., \&
  {Friedjung}, M. 2000, \aaps, 146, 407

\bibitem[{{Bessell} \& {Brett}(1988)}]{1988PASP..100.1134B}
{Bessell}, M.~S. \& {Brett}, J.~M. 1988, \pasp, 100, 1134

\bibitem[{{Corradi} {et~al.}(2008){Corradi}, {Rodr{\'{\i}}guez-Flores},
  {Mampaso}, {Greimel}, {Viironen}, {Drew}, {Lennon}, {Mikolajewska}, {Sabin},
  \& {Sokoloski}}]{2008A&A...480..409C}
{Corradi}, R.~L.~M., {Rodr{\'{\i}}guez-Flores}, E.~R., {Mampaso}, A., {et~al.}
  2008, \aap, 480, 409

\bibitem[{{Cross} {et~al.}(2012){Cross}, {Collins}, {Mann}, {Read}, {Sutorius},
  {Blake}, {Holliman}, {Hambly}, {Emerson}, {Lawrence}, \&
  {Noddle}}]{2012A&A...548A.119C}
{Cross}, N.~J.~G., {Collins}, R.~S., {Mann}, R.~G., {et~al.} 2012, \aap, 548,
  A119

\bibitem[{{Dalton} {et~al.}(2006){Dalton}, {Caldwell}, {Ward}, {Whalley},
  {Woodhouse}, {Edeson}, {Clark}, {Beard}, {Gallie}, {Todd}, {Strachan},
  {Bezawada}, {Sutherland}, \& {Emerson}}]{2006SPIE.6269E..30D}
{Dalton}, G.~B., {Caldwell}, M., {Ward}, A.~K., {et~al.} 2006, {in Society of
  Photo-Optical Instrumentation Engineers (SPIE) Conference Series, Vol. 6269,
  Society of Photo-Optical Instrumentation Engineers (SPIE) Conference Series}

\bibitem[{{Downes} {et~al.}(2001){Downes}, {Webbink}, {Shara}, {Ritter},
  {Kolb}, \& {Duerbeck}}]{2001PASP..113..764D}
{Downes}, R.~A., {Webbink}, R.~F., {Shara}, M.~M., {et~al.} 2001, \pasp, 113,
  764

\bibitem[{{Drew} {et~al.}(2005){Drew}, {Greimel}, {Irwin}, {Aungwerojwit},
  {Barlow}, {Corradi}, {Drake}, {G{\"a}nsicke}, {Groot}, {Hales}, {Hopewell},
  {Irwin}, {Knigge}, {Leisy}, {Lennon}, {Mampaso}, {Masheder}, {Matsuura},
  {Morales-Rueda}, {Morris}, {Parker}, {Phillipps}, {Rodriguez-Gil}, {Roelofs},
  {Skillen}, {Sokoloski}, {Steeghs}, {Unruh}, {Viironen}, {Vink}, {Walton},
  {Witham}, {Wright}, {Zijlstra}, \& {Zurita}}]{2005MNRAS.362..753D}
{Drew}, J.~E., {Greimel}, R., {Irwin}, M.~J., {et~al.} 2005, \mnras, 362, 753

\bibitem[{{Emerson} {et~al.}(2006){Emerson}, {McPherson}, \&
  {Sutherland}}]{2006Msngr.126...41E}
{Emerson}, J., {McPherson}, A., \& {Sutherland}, W. 2006, The Messenger, 126,
  41

\bibitem[{{Emerson} \& {Sutherland}(2010)}]{2010Msngr.139....2E}
{Emerson}, J. \& {Sutherland}, W. 2010, The Messenger, 139, 2

\bibitem[{{Epchtein} {et~al.}(1997){Epchtein}, {de Batz}, {Capoani},
  {Chevallier}, {Copet}, {Fouqu{\'e}}, {Lacombe}, {Le Bertre}, {Pau}, {Rouan},
  {Ruphy}, {Simon}, {Tiph{\`e}ne}, {Burton}, {Bertin}, {Deul}, {Habing},
  {Borsenberger}, {Dennefeld}, {Guglielmo}, {Loup}, {Mamon}, {Ng}, {Omont},
  {Provost}, {Renault}, {Tanguy}, {Kimeswenger}, {Kienel}, {Garzon}, {Persi},
  {Ferrari-Toniolo}, {Robin}, {Paturel}, {Vauglin}, {Forveille}, {Delfosse},
  {Hron}, {Schultheis}, {Appenzeller}, {Wagner}, {Balazs}, {Holl},
  {L{\'e}pine}, {Boscolo}, {Picazzio}, {Duc}, \&
  {Mennessier}}]{1997Msngr..87...27E}
{Epchtein}, N., {de Batz}, B., {Capoani}, L., {et~al.} 1997, The Messenger, 87,
  27

\bibitem[{{Ferland} {et~al.}(1998){Ferland}, {Korista}, {Verner}, {Ferguson},
  {Kingdon}, \& {Verner}}]{1998PASP..110..761F}
{Ferland}, G.~J., {Korista}, K.~T., {Verner}, D.~A., {et~al.} 1998, \pasp, 110,
  761

\bibitem[{{Garcia-Lario} {et~al.}(1997){Garcia-Lario}, {Manchado}, {Pych}, \&
  {Pottasch}}]{1997A&AS..126..479G}
{Garcia-Lario}, P., {Manchado}, A., {Pych}, W., \& {Pottasch}, S.~R. 1997,
  \aaps, 126, 479

\bibitem[{{Gonzalez} {et~al.}(2011){Gonzalez}, {Rejkuba}, {Zoccali}, {Valenti},
  \& {Minniti}}]{2011A&A...534A...3G}
{Gonzalez}, O.~A., {Rejkuba}, M., {Zoccali}, M., {Valenti}, E., \& {Minniti},
  D. 2011, \aap, 534, A3

\bibitem[{{Hewett} {et~al.}(2006){Hewett}, {Warren}, {Leggett}, \&
  {Hodgkin}}]{2006MNRAS.367..454H}
{Hewett}, P.~C., {Warren}, S.~J., {Leggett}, S.~K., \& {Hodgkin}, S.~T. 2006,
  \mnras, 367, 454

\bibitem[{{Hyung} {et~al.}(2001){Hyung}, {Aller}, {Feibelman}, \&
  {Lee}}]{2001ApJ...563..889H}
{Hyung}, S., {Aller}, L.~H., {Feibelman}, W.~A., \& {Lee}, S.-J. 2001, \apj,
  563, 889

\bibitem[{{Irwin} {et~al.}(2004){Irwin}, {Lewis}, {Hodgkin}, {Bunclark},
  {Evans}, {McMahon}, {Emerson}, {Stewart}, \& {Beard}}]{2004SPIE.5493..411I}
{Irwin}, M.~J., {Lewis}, J., {Hodgkin}, S., {et~al.} 2004, in Society of
  Photo-Optical Instrumentation Engineers (SPIE) Conference Series, Vol. 5493,
  Society of Photo-Optical Instrumentation Engineers (SPIE) Conference Series,
  ed. {P.~J.~Quinn \& A.~Bridger}, 411--422

\bibitem[{{Jacoby} \& {Van de Steene}(2004)}]{2004A&A...419..563J}
{Jacoby}, G.~H. \& {Van de Steene}, G. 2004, \aap, 419, 563

\bibitem[{{Kerber} {et~al.}(2003){Kerber}, {Mignani}, {Guglielmetti}, \&
  {Wicenec}}]{2003A&A...408.1029K}
{Kerber}, F., {Mignani}, R.~P., {Guglielmetti}, F., \& {Wicenec}, A. 2003,
  \aap, 408, 1029

\bibitem[{{Kharchenko} {et~al.}(2002){Kharchenko}, {Kilpio}, {Malkov}, \&
  {Schilbach}}]{2002A&A...384..925K}
{Kharchenko}, N., {Kilpio}, E., {Malkov}, O., \& {Schilbach}, E. 2002, \aap,
  384, 925

\bibitem[{{Kinman} {et~al.}(1988){Kinman}, {Feast}, \&
  {Lasker}}]{1988AJ.....95..804K}
{Kinman}, T.~D., {Feast}, M.~W., \& {Lasker}, B.~M. 1988, \aj, 95, 804

\bibitem[{{Lewis}(2006)}]{2006IAUS..234..449L}
{Lewis}, B.~M. 2006, in IAU Symposium, Vol. 234, Planetary Nebulae in our
  Galaxy and Beyond, ed. {M.~J.~Barlow \& R.~H.~M{\'e}ndez}, 449--450

\bibitem[{{Maciel} \& {Costa}(2003)}]{2003IAUS..209..551M}
{Maciel}, W.~J. \& {Costa}, R.~D.~D. 2003, in IAU Symposium, Vol. 209,
  Planetary Nebulae: Their Evolution and Role in the Universe, ed. {S.~Kwok,
  M.~Dopita, \& R.~Sutherland}, 551

\bibitem[{{Martins} {et~al.}(2005){Martins}, {Schaerer}, \&
  {Hillier}}]{2005A&A...436.1049M}
{Martins}, F., {Schaerer}, D., \& {Hillier}, D.~J. 2005, \aap, 436, 1049

\bibitem[{{Minniti} {et~al.}(2010){Minniti}, {Lucas}, {Emerson}, {Saito},
  {Hempel}, {Pietrukowicz}, {Ahumada}, {Alonso}, {Alonso-Garcia}, {Arias},
  {Bandyopadhyay}, {Barb{\'a}}, {Barbuy}, {Bedin}, {Bica}, {Borissova},
  {Bronfman}, {Carraro}, {Catelan}, {Clari{\'a}}, {Cross}, {de Grijs},
  {D{\'e}k{\'a}ny}, {Drew}, {Fari{\~n}a}, {Feinstein}, {Fern{\'a}ndez
  Laj{\'u}s}, {Gamen}, {Geisler}, {Gieren}, {Goldman}, {Gonzalez}, {Gunthardt},
  {Gurovich}, {Hambly}, {Irwin}, {Ivanov}, {Jord{\'a}n}, {Kerins}, {Kinemuchi},
  {Kurtev}, {L{\'o}pez-Corredoira}, {Maccarone}, {Masetti}, {Merlo},
  {Messineo}, {Mirabel}, {Monaco}, {Morelli}, {Padilla}, {Palma}, {Parisi},
  {Pignata}, {Rejkuba}, {Roman-Lopes}, {Sale}, {Schreiber}, {Schr{\"o}der},
  {Smith}, {Sodr{\'e}}, {Soto}, {Tamura}, {Tappert}, {Thompson}, {Toledo},
  {Zoccali}, \& {Pietrzynski}}]{2010NewA...15..433M}
{Minniti}, D., {Lucas}, P.~W., {Emerson}, J.~P., {et~al.} 2010, \na, 15, 433

\bibitem[{{Miszalski} {et~al.}(2009){Miszalski}, {Acker}, {Moffat}, {Parker},
  \& {Udalski}}]{2009A&A...496..813M}
{Miszalski}, B., {Acker}, A., {Moffat}, A.~F.~J., {Parker}, Q.~A., \&
  {Udalski}, A. 2009, \aap, 496, 813

\bibitem[{{Miszalski} {et~al.}(2011){Miszalski}, {Napiwotzki}, {Cioni},
  {Groenewegen}, {Oliveira}, \& {Udalski}}]{2011A&A...531A.157M}
{Miszalski}, B., {Napiwotzki}, R., {Cioni}, M.-R.~L., {et~al.} 2011, \aap, 531,
  A157

\bibitem[{{Miszalski} {et~al.}(2008){Miszalski}, {Parker}, {Acker}, {Birkby},
  {Frew}, \& {Kovacevic}}]{2008MNRAS.384..525M}
{Miszalski}, B., {Parker}, Q.~A., {Acker}, A., {et~al.} 2008, \mnras, 384, 525

\bibitem[{{Parker} {et~al.}(2006){Parker}, {Acker}, {Frew}, {Hartley},
  {Peyaud}, {Ochsenbein}, {Phillipps}, {Russeil}, {Beaulieu}, {Cohen},
  {K{\"o}ppen}, {Miszalski}, {Morgan}, {Morris}, {Pierce}, \&
  {Vaughan}}]{2006MNRAS.373...79P}
{Parker}, Q.~A., {Acker}, A., {Frew}, D.~J., {et~al.} 2006, \mnras, 373, 79

\bibitem[{{Parker} {et~al.}(2005){Parker}, {Phillipps}, {Pierce}, {Hartley},
  {Hambly}, {Read}, {MacGillivray}, {Tritton}, {Cass}, {Cannon}, {Cohen},
  {Drew}, {Frew}, {Hopewell}, {Mader}, {Malin}, {Masheder}, {Morgan}, {Morris},
  {Russeil}, {Russell}, \& {Walker}}]{2005MNRAS.362..689P}
{Parker}, Q.~A., {Phillipps}, S., {Pierce}, M.~J., {et~al.} 2005, \mnras, 362,
  689

\bibitem[{{Pena} \& {Torres-Peimbert}(1987)}]{1987RMxAA..14..534P}
{Pena}, M. \& {Torres-Peimbert}, S. 1987, \rmxaa, 14, 534

\bibitem[{{Petrosian}(1976)}]{1976ApJ...209L...1P}
{Petrosian}, V. 1976, \apjl, 209, L1

\bibitem[{{Phillips} \& {Cuesta}(1994)}]{1994A&AS..104..169P}
{Phillips}, J.~P. \& {Cuesta}, L. 1994, \aaps, 104, 169

\bibitem[{{Phillips} \& {Zepeda-Garc{\'{\i}}a}(2009)}]{2009MNRAS.394.1875P}
{Phillips}, J.~P. \& {Zepeda-Garc{\'{\i}}a}, D. 2009, \mnras, 394, 1875

\bibitem[{{Pottasch}(1984)}]{1984ASSL..107.....P}
{Pottasch}, S.~R., ed. 1984, Astrophysics and Space Science Library, Vol. 107,
  {Planetary nebulae - A study of late stages of stellar evolution}

\bibitem[{{Ramos-Larios} {et~al.}(2009){Ramos-Larios}, {Guerrero},
  {Su{\'a}rez}, {Miranda}, \& {G{\'o}mez}}]{2009A&A...501.1207R}
{Ramos-Larios}, G., {Guerrero}, M.~A., {Su{\'a}rez}, O., {Miranda}, L.~F., \&
  {G{\'o}mez}, J.~F. 2009, \aap, 501, 1207

\bibitem[{{Ramos-Larios} \& {Phillips}(2005)}]{2005MNRAS.357..732R}
{Ramos-Larios}, G. \& {Phillips}, J.~P. 2005, \mnras, 357, 732

\bibitem[{{Rauch}(2003)}]{2003A&A...403..709R}
{Rauch}, T. 2003, \aap, 403, 709

\bibitem[{{Rudy} {et~al.}(2001){Rudy}, {Lynch}, {Mazuk}, {Puetter}, \&
  {Dearborn}}]{2001AJ....121..362R}
{Rudy}, R.~J., {Lynch}, D.~K., {Mazuk}, S., {Puetter}, R.~C., \& {Dearborn},
  D.~S.~P. 2001, \aj, 121, 362

\bibitem[{{Ruffle} {et~al.}(2004){Ruffle}, {Zijlstra}, {Walsh}, {Gray},
  {Gesicki}, {Minniti}, \& {Comeron}}]{2004MNRAS.353..796R}
{Ruffle}, P.~M.~E., {Zijlstra}, A.~A., {Walsh}, J.~R., {et~al.} 2004, \mnras,
  353, 796

\bibitem[{{Saito} \& {et al.}(2012)}]{2012A&A...537A.107S}
{Saito}, R.~K. \& {et al.} 2012, \aap, 537, A107

\bibitem[{{Skrutskie} {et~al.}(2006){Skrutskie}, {Cutri}, {Stiening},
  {Weinberg}, {Schneider}, {Carpenter}, {Beichman}, {Capps}, {Chester},
  {Elias}, {Huchra}, {Liebert}, {Lonsdale}, {Monet}, {Price}, {Seitzer},
  {Jarrett}, {Kirkpatrick}, {Gizis}, {Howard}, {Evans}, {Fowler}, {Fullmer},
  {Hurt}, {Light}, {Kopan}, {Marsh}, {McCallon}, {Tam}, {Van Dyk}, \&
  {Wheelock}}]{2006AJ....131.1163S}
{Skrutskie}, M.~F., {Cutri}, R.~M., {Stiening}, R., {et~al.} 2006, \aj, 131,
  1163

\bibitem[{{Tylenda} {et~al.}(1992){Tylenda}, {Acker}, {Stenholm}, \&
  {Koeppen}}]{1992A&AS...95..337T}
{Tylenda}, R., {Acker}, A., {Stenholm}, B., \& {Koeppen}, J. 1992, \aaps, 95,
  337

\bibitem[{{van Hoof}(2000)}]{2000MNRAS.314...99V}
{van Hoof}, P.~A.~M. 2000, \mnras, 314, 99

\bibitem[{{van Hoof} {et~al.}(2010){van Hoof}, {van de Steene}, {Barlow},
  {Exter}, {Sibthorpe}, {Ueta}, {Peris}, {Groenewegen}, {Blommaert}, {Cohen},
  {De Meester}, {Ferland}, {Gear}, {Gomez}, {Hargrave}, {Huygen}, {Ivison},
  {Jean}, {Leeks}, {Lim}, {Olofsson}, {Polehampton}, {Regibo}, {Royer},
  {Swinyard}, {Vandenbussche}, {van Winckel}, {Waelkens}, {Walker}, \&
  {Wesson}}]{2010A&A...518L.137V}
{van Hoof}, P.~A.~M., {van de Steene}, G.~C., {Barlow}, M.~J., {et~al.} 2010,
  \aap, 518, L137

\bibitem[{{Viironen} {et~al.}(2009{\natexlab{a}}){Viironen}, {Greimel},
  {Corradi}, {Mampaso}, {Rodr{\'{\i}}guez}, {Sabin}, {Delgado-Inglada}, {Drew},
  {Giammanco}, {Gonz{\'a}lez-Solares}, {Irwin}, {Miszalski}, {Parker},
  {Rodr{\'{\i}}guez-Flores}, \& {Zijlstra}}]{2009A&A...504..291V}
{Viironen}, K., {Greimel}, R., {Corradi}, R.~L.~M., {et~al.}
  2009{\natexlab{a}}, \aap, 504, 291

\bibitem[{{Viironen} {et~al.}(2009{\natexlab{b}}){Viironen}, {Mampaso},
  {Corradi}, {Rodr{\'{\i}}guez}, {Greimel}, {Sabin}, {Sale}, {Unruh},
  {Delgado-Inglada}, {Drew}, {Giammanco}, {Groot}, {Parker}, {Sokoloski}, \&
  {Zijlstra}}]{2009A&A...502..113V}
{Viironen}, K., {Mampaso}, A., {Corradi}, R.~L.~M., {et~al.}
  2009{\natexlab{b}}, \aap, 502, 113

\bibitem[{{Whitelock}(1985)}]{1985MNRAS.213...59W}
{Whitelock}, P.~A. 1985, \mnras, 213, 59

\bibitem[{{Yasuda} {et~al.}(2001){Yasuda}, {Fukugita}, {Narayanan}, {Lupton},
  {Strateva}, {Strauss}, {Ivezi{\'c}}, {Kim}, {Hogg}, {Weinberg}, {Shimasaku},
  {Loveday}, {Annis}, {Bahcall}, {Blanton}, {Brinkmann}, {Brunner}, {Connolly},
  {Csabai}, {Doi}, {Hamabe}, {Ichikawa}, {Ichikawa}, {Johnston}, {Knapp},
  {Kunszt}, {Lamb}, {McKay}, {Munn}, {Nichol}, {Okamura}, {Schneider},
  {Szokoly}, {Vogeley}, {Watanabe}, \& {York}}]{2001AJ....122.1104Y}
{Yasuda}, N., {Fukugita}, M., {Narayanan}, V.~K., {et~al.} 2001, \aj, 122, 1104

\bibitem[{{Zhang} {et~al.}(2005){Zhang}, {Chen}, \&
  {Yang}}]{2005NewA...10..325Z}
{Zhang}, P., {Chen}, P.~S., \& {Yang}, H.~T. 2005, \na, 10, 325

\bibitem[{{Zijlstra} {et~al.}(2008){Zijlstra}, {van Hoof}, \&
  {Perley}}]{2008ApJ...681.1296Z}
{Zijlstra}, A.~A., {van Hoof}, P.~A.~M., \& {Perley}, R.~A. 2008, \apj, 681,
  1296

\end{thebibliography}
%\end{thebibliography}

%%%%%%%%%%%%%%%%%%%%%%%%%%%%%%%%%%%%%%%%%%%%%%%%%%%%%%%%%%%%%%%%%%%%%%%%%%%%%%%%%%%%%%%%%%%%%%%%
%%%%%%%%%%%%%%%%%%%%%%%%%%%%%%%%%%%%%%%%%%%%%%%%%%%%%%%%%%%%%%%%%%%%%%%%%%%%%%%%%%%%%%%%%%%%%%%%
%%%%%%%%%%%%%%%%%%%%%%%%%%%%%%%%%%%%%%%%%%%%%%%%%%%%%%%%%%%%%%%%%%%%%%%%%%%%%%%%%%%%%%%%%%%%%%%%
%%%%%%%%%%%%%%%%%%%%%%%%%%%%%%%%%%%%%%%%%%%%%%%%%%%%%%%%%%%%%%%%%%%%%%%%%%%%%%%%%%%%%%%%%%%%%%%%
%%%%%%%%%%%%%%%%%%%%%%%%%%%%%%%%%%%%%%%%%%%%%%%%%%%%%%%%%%%%%%%%%%%%%%%%%%%%%%%%%%%%%%%%%%%%%%%%
%%%%%%%%%%%%%%%%%%%%%%%%%%%%%%%%%%%%%%%%%%%%%%%%%%%%%%%%%%%%%%%%%%%%%%%%%%%%%%%%%%%%%%%%%%%%%%%%

\longtabL{3}{
\begin{landscape}
\begin{longtable}{lcccrcccccccccc}
\caption{\label{tablita} NIR photometry catalogue of PNe (aperture $1.41\arcsec$).
The marks ``:" and ``*" refer to near saturation
and probable not central star of PN, respectively.
While the mark ``$\dagger$" refer to the new NIR determination.}\\
\hline\hline
name & PN G & $\alpha$(2000) & $\delta$(2000) & OT & $Z$ & $\sigma$$Z$ & $Y$ & $\sigma$$Y$ & $J$ & $\sigma$$J$ & $H$ & $\sigma$$H$ & $K_{\rm  s}$ & $\sigma$$K_{\rm  s}$  \\
\hline
\endfirsthead
\caption{continued.}\\
\hline\hline
name & PN G & $\alpha$(2000) & $\delta$(2000) & OT & $Z$ & $\sigma$$Z$ & $Y$ & $\sigma$$Y$ & $J$ & $\sigma$$J$ & $H$ & $\sigma$$H$ & $K_{\rm  s}$ & $\sigma$$K_{\rm  s}$  \\
\hline
\endhead
\hline
\endfoot
JaSt 71 $\dagger$ &  000.0$-$01.2   &  17 50 23.21  &  $-$28 33 10.65  &  -1  &  -  &  -  &  -  &  -  &  -  &  -  &  -  &  -  &  12.888  &  0.026  \\
H 1-62          &  000.0$-$06.8  &  18 13 17.96  &  $-$32 19 42.92  &  1  &  -  &  -  &  -  &  -  &  13.005  &  0.002  &  13.041  &  0.004  &  12.456  &  0.004  \\
JaSt 69 $\dagger$ &  000.1$-$01.0   &  17 50 10.01  &  $-$29 19 04.28  &  1  &  -  &  -  &  -  &  -  &  17.833  &  0.221  &  15.387  &  0.095  &  14.236  &  0.072  \\
M 3-43          &  000.1$-$01.1  &  17 50 24.34  &  $-$29 25 18.46  &  1  &  -  &  -  &  -  &  -  &  13.684  &  0.005  &  12.96  &  0.011  &  -  &  -  \\
JaSt 75  &  000.1$-$01.2   &  17 50 47.99  &  $-$29 24 43.43  &  1  &  -  &  -  &  -  &  -  &  14.407  &  0.01  &  13.738  &  0.021  &  13.038  &  0.024  \\
H 2-40 $\dagger$         &  000.1$-$05.6  &  18 08 30.76  &  $-$31 36 35.12  &  1  &  -  &  -  &  -  &  -  &  15.764  &  0.035  &  15.305  &  0.043  &  15.13  &  0.06  \\
Ae 2-J $\dagger$         &  000.1+02.6  &  17 35 35.45  &  $-$27 24 05.31  &  -1  &  15.607  &  0.017  &  15.044  &  0.018  &  -  &  -  &  -  &  -  &  -  &  -  \\
H 1-16          &  000.1+04.3  &  17 29 23.39  &  $-$26 26 05.11  &  1  &  -  &  -  &  -  &  -  &  13.055  &  0.003  &  12.603  &  0.004  &  11.749  &  0.003  \\
\end{longtable}
\end{landscape}
}% End \longtabL

%%%%%%%%%%%%%%%%%%%%%%%%%%%%%%%%%%%%%%%%%%%%%%%%%%%%%%%%%%%%%%%%%%%%%%%%%%%%%%%%
%%%%%%%%%%%%%%%%%%%%%%%%%%%%%%%%%%%%%%%%%%%%%%%%%%%%%%%%%%%%%%%%%%%%%%%%%%%%%%%%
%%%%%%%%%%%%%%%%%%%%%%%%%%%%%%%%%%%%%%%%%%%%%%%%%%%%%%%%%%%%%%%%%%%%%%%%%%%%%%%%
%%%%%%%%%%%%%%%%%%%%%%%%%%%%%%%%%%%%%%%%%%%%%%%%%%%%%%%%%%%%%%%%%%%%%%%%%%%%%%%%
%%%%%%%%%%%%%%%%%%%%%%%%%%%%%%%%%%%%%%%%%%%%%%%%%%%%%%%%%%%%%%%%%%%%%%%%%%%%%%%%
%%%%%%%%%%%%%%%%%%%%%%%%%%%%%%%%%%%%%%%%%%%%%%%%%%%%%%%%%%%%%%%%%%%%%%%%%%%%%%%%
%%%%%%%%%%%%%%%%%%%%%%%%%%%%%%%%%%%%%%%%%%%%%%%%%%%%%%%%%%%%%%%%%%%%%%%%%%%%%%%%

\longtabL{4}{
\begin{landscape}
\begin{longtable}{lcccccccccc}
\caption{\label{tablita-3} Petrosian radius of the 353 PNe listed in Table~\ref{tablita}.
When $\sigma$=0 means that $\sigma$$<0.05$ and when $\sigma$$\equiv n$ 
means that there is only one measured so it is not possible to give
an stdev.}\\
\hline\hline
name & P$_Z$ & $\sigma$ & P$_Y$ & $\sigma$ & P$_J$ & $\sigma$ & P$_H$ & $\sigma$ & P$_{Ks}$ & $\sigma$ \\
\hline
\endfirsthead
\caption{continued.}\\
\hline\hline
name & P$_Z$ & $\sigma$ & P$_Y$ & $\sigma$ & P$_J$ & $\sigma$ & P$_H$ & $\sigma$ & P$_{Ks}$ & $\sigma$ \\
\hline
\endhead
\hline
\endfoot
Ae 1  &  -  &  -  &  -  &  -  &  2.2  &  0.7  &  1.2  &  0.1  &  1.5  &  0.2  \\
Ae 2-E  &  3.0  &  1.2  &  2.9  &  0.4  &  -  &  -  &  -  &  -  &  -  &  -  \\
Ae 2-F  &  2.8  &  0.8  &  1.7  &  0.1  &  1.7  &  n  &  -  &  -  &  1.8  &  n  \\
Ae 2-G  &  1.7  &  0.2  &  1.6  &  0.0  &  -  &  -  &  -  &  -  &  -  &  -  \\
Ae 2-H  &  -  &  -  &  -  &  -  &  2.2  &  n  &  1.7  &  n  &  1.7  &  0.1  \\
Ae 2-I  &  3.2  &  0.0  &  3.1  &  0.1  &  -  &  -  &  -  &  -  &  -  &  -  \\
Ae 2-J  &  2.9  &  0.0  &  2.5  &  0.0  &  -  &  -  &  -  &  -  &  -  &  -  \\
Ae 2-K  &  -  &  -  &  3.3  &  0.0  &  -  &  -  &  -  &  -  &  -  &  -  \\
\end{longtable}
\end{landscape}
}% End \longtabL

%%%%%%%%%%%%%%%%%%%%%%%%%%%%%%%%%%%%%%%%%%%%%%%%%%%%%%%%%%%%%%%%%%%%%%%%%%%%%%%%
%%%%%%%%%%%%%%%%%%%%%%%%%%%%%%%%%%%%%%%%%%%%%%%%%%%%%%%%%%%%%%%%%%%%%%%%%%%%%%%%
%%%%%%%%%%%%%%%%%%%%%%%%%%%%%%%%%%%%%%%%%%%%%%%%%%%%%%%%%%%%%%%%%%%%%%%%%%%%%%%%
%%%%%%%%%%%%%%%%%%%%%%%%%%%%%%%%%%%%%%%%%%%%%%%%%%%%%%%%%%%%%%%%%%%%%%%%%%%%%%%%
%%%%%%%%%%%%%%%%%%%%%%%%%%%%%%%%%%%%%%%%%%%%%%%%%%%%%%%%%%%%%%%%%%%%%%%%%%%%%%%%
%%%%%%%%%%%%%%%%%%%%%%%%%%%%%%%%%%%%%%%%%%%%%%%%%%%%%%%%%%%%%%%%%%%%%%%%%%%%%%%%

\longtabL{5}{
\begin{landscape}
\begin{longtable}{lcccc}
\caption{\label{tablita-2} Planetary nebulae that were not measured.}\\
\hline\hline
name & PN G & $\alpha$(2000) & $\delta$(2000) & status  \\
\hline
\endfirsthead
\caption{continued.}\\
\hline\hline
name & PN G & $\alpha$(2000) & $\delta$(2000) & status  \\
\hline
\endhead
\hline
\endfoot
JaSt 93        & 000.1$-$01.9  & 17 53 24.43  & $-$29 49 47.7 & N/D \\
Bl 3-10        & 000.1$-$02.3 & 17 55 20.48  & $-$29 57 33.4  & N/D \\
K 6-8          & 000.2+01.7  & 17 39 39.253  & $-$27 47 23.56  & N/D \\
PHR1752-2930 & 000.3$-$01.6 & 17 52 52.20  & $-$29 30 00.1  & N/D \\
JaSt 86        & 000.3$-$01.6  & 17 52 52.20  & $-$29 30 00.1 & N/D \\
Sa 3-117       & 000.3$-$02.8 & 17 57 43.367  & $-$30 02 29.91  & N/D \\
M 3-47         & 000.3$-$04.6 & 18 05 02.695  & $-$30 58 17.39  & ext \\
M 2-20         & 000.4$-$02.9 & 17 58 19.337  & $-$30 00 39.32  & N/D \\
JaSt 96  & 000.5$-$01.7  & 17 25 03.470  & $-$31 28 38.50 & N/D \\
Ae 2-Q         & 000.5$-$03.1 & 17 59 15.585  & $-$30 02 47.15 & N/D \\
SB 2           & 000.5$-$05.3  & 18 08 34.7  & $-$31 06 52  & N/D \\
JaSt 77        & 000.6$-$01.0  & 17 51 11.504  & $-$28 56 26.16 & N/D \\
KnFs 1         & 000.6$-$01.3 & 17 52 35.87  & $-$29 06 38.6 & N/D \\
SB 3           & 000.7$-$06.1  & 18 12 14.48  & $-$31 19 58.0  & N/D \\
M 2-35         & 000.7$-$07.4 & 18 17 37.195  & $-$31 56 46.86  & ext \\
JaSt 38 & 000.8+01.3  & 17 42 32.674  & $-$27 33 08.82  & N/D \\
JaSt 44 & 000.9+01.1  & 17 43 23.305  & $-$27 34 03.57 & N/D \\
PHR1804-3016 & 000.9$-$04.2 & 18 04 48.1  & $-$30 16 49  & N/D \\
M 3-23         & 000.9$-$04.8 & 18 07 06.148  & $-$30 34 16.96  & ext \\
JaSt 41        & 001.0+01.3  & 17 42 49.96  & $-$27 21 19.7 & N/D \\
K 1-4          & 001.0+01.9 & 17 40 27.366  & $-$27 01 02.91  & N/D \\
JaSt 89 & 001.3$-$01.0  & 17 53 06.57  & $-$28 18 09.2 & N/D \\
SAWI 1         & 001.4$-$03.4 & 18 02 25.85  & $-$29 25 05.4 & N/D \\
K 6-17         & 001.5$-$00.7  & 17 52 08.717  & $-$28 02 16.56  & N/D \\
JaSt 55        & 001.6+00.9  & 17 45 37.36  & $-$27 01 18.4 & N/D \\
K 6-10         & 001.6+01.5  & 17 43 16.950  & $-$26 44 17.54  & N/D \\
Bl Q           & 001.6$-$01.3 & 17 54 34.94  & $-$28 12 43.3  & N/D \\
SB 6           & 001.6$-$05.9  & 18 13 15.842  & $-$30 25 58.49  & N/D \\
H 1-45         & 002.0$-$02.0 & 17 58 21.87  & $-$28 14 52.3  & ext \\
KnFs 2         & 002.2$-$02.5 & 18 00 59.92  & $-$28 16 19.8  & N/D \\
PHR1808-2913 & 002.2$-$04.3 & 18 08 06.42  & $-$29 13 15.7 & N/D \\
Cn 1-5         & 002.2$-$09.4 & 18 29 11.675  & $-$31 29 58.81  & ext \\
PHR1744-2603 & 002.3+01.7 & 17 44 35.4  & $-$26 03 36  & N/D \\
Ta 5           & 002.3+02.2 & 17 42 30.10  & $-$25 45 28.7  & N/D \\
PPA1741-2538 & 002.3+02.4 & 17 41 48.4  & $-$25 38 18  & N/D \\
Sa 3-115       & 002.4$-$03.2 & 18 04 05.438  & $-$28 27 50.90  & N/D \\
Pe 2-11        & 002.5$-$01.7 & 17 58 31.27  & $-$27 37 05.8  & S/D \\
Ta 1580        & 002.6+02.1 & 17 43 39.442  & $-$25 36 42.51  & N/D \\
PPA1745-2542 & 002.7+01.7 & 17 45 18.9  & $-$25 42 05  & N/D \\
M 1-42         & 002.7$-$04.8 & 18 11 05.07  & $-$28 58 58.5  & ext \\
Pe 2-12        & 002.8$-$02.2 & 18 01 10.300  & $-$27 38 19.88  & S/D \\
Hb 4           & 003.1+02.9 & 17 41 52.763  & $-$24 42 08.07 & ext \\
K 5-7          & 003.1+04.1  & 17 37 20.167  & $-$24 03 27.84  & N/D \\
SB 7           & 003.3$-$06.1  & 18 17 46.363  & $-$29 06 07.52 & N/D \\
MPA1748-2511 & 003.5+01.3 & 17 48 41.629  & $-$25 11 33.37  & ext \\
NGC 6565       & 003.5$-$04.6 & 18 11 52.470  & $-$28 10 42.27 & ext \\
PHR1759-2630 & 003.6$-$01.3 & 17 59 12.09  & $-$26 30 24.8  & ext \\
Ta 2111        & 003.9+01.6 & 17 48 28.465  & $-$24 41 25.07  & N/D \\
M 1-35         & 003.9$-$02.3 & 18 03 39.30  & $-$26 43 33.9 & ext \\
KnFs 7         & 003.9$-$03.1 & 18 06 50.05  & $-$27 06 16.09  & S/D \\
M 2-29         & 004.0$-$03.0 & 18 06 40.862  & $-$26 54 55.95  & S/D \\
Pe 1-12        & 004.0$-$05.8 & 18 17 42.37  & $-$28 17 16.5  & N/D \\
KnFs 10        & 004.2$-$03.2 & 18 08 01.312  & $-$26 54 01.52 & S/D \\
SB 8           & 004.2$-$05.2  & 18 15 50.3  & $-$27 49 00  & N/D \\
H 1-53         & 004.3$-$02.6 & 18 05 57.43  & $-$26 29 42.0 & S/D \\
MPA1748-2402 & 004.5+02.0 & 17 48 21.26  & $-$24 02 14.0  & N/D \\
SB 10          & 004.7$-$05.5  & 18 18 06.9  & $-$27 31 35  & N/D \\
M 1-25         & 004.9+04.9 & 17 38 30.32  & $-$22 08 38.8 & ext \\
H 1-27         & 005.0+04.4 & 17 40 17.95  & $-$22 19 18.0  & S/D \\
M 3-13         & 005.2+04.2 & 17 41 36.62  & $-$22 13 03.0 & S/D \\
SB 11          & 005.2$-$05.9  & 18 20 44.7  & $-$27 15 48  & N/D \\
M 3-24         & 005.5$-$02.5 & 18 07 53.914  & $-$25 24 02.71  & ext \\
H 2-44         & 005.5$-$04.0 & 18 13 40.45  & $-$26 08 38.3 & N/D \\
NGC 6620       & 005.8$-$06.1 & 18 22 54.17  & $-$26 49 17.1  & ext \\
KnFs 15        & 006.2$-$03.7 & 18 14 19.324  & $-$25 20 51.22  & N/D \\
H 2-18         & 006.3+04.4 & 17 43 28.751  & $-$21 09 51.29  & S/D \\
Pe 2-13        & 006.4$-$04.6 & 18 18 13.41  & $-$25 38 10.4 & N/D \\
M 1-41         & 006.7$-$02.2 & 18 09 29.902  & $-$24 12 23.46  & N/D \\
Th 4-7         & 006.8+02.3 & 17 52 22.569  & $-$21 51 13.43  & ext \\
M 3-15         & 006.8+04.1 & 17 45 31.74  & $-$20 58 01.8 & ext \\
Hb 6           & 007.2+01.8 & 17 55 07.023  & $-$21 44 39.98  & ext \\
Th 4-1         & 007.5+04.3 & 17 46 20.82  & $-$20 13 48.3 & S/D \\
H 1-65         & 007.8$-$04.4 & 18 20 08.85  & $-$24 15 05.0 & S/D \\
NGC 6445       & 008.0+03.9 & 17 49 15.21  & $-$20 00 34.5  & ext \\
M 1-40         & 008.3$-$01.1 & 18 08 25.994  & $-$22 16 53.25  & ext \\
He 2-406       & 008.6$-$07.0 & 18 31 52.845  & $-$24 46 16.55  & ext \\
PHR1751-1925 & 008.8+03.8 & 17 51 08.7  & $-$19 25 47  & N/D \\
NGC 6629       & 009.4$-$05.0 & 18 25 42.458  & $-$23 12 10.23  & ext \\
M 3-32         & 009.4$-$09.8 & 18 44 43.03  & $-$25 21 34.8  & ext \\
H 1-67         & 009.8$-$04.6 & 18 25 04.976  & $-$22 34 52.64  & ext \\
IRAS18333-2357 & 009.8$-$07.5 & 18 36 22.85  & $-$23 55 19.5  & N/D \\
BMP1145-6328 & 295.6$-$01.5 & 11 45 09.4 & $-$63 28 54 & N/D \\
PHR1206-6122 & 297.5+01.0 & 12 06 25.5 & $-$61 22 44 & N/D \\
PHR1203-6403 & 297.6$-$01.6 & 12 03 12.4 & $-$64 03 41 & N/D \\
PHR1212-6043 & 298.2+01.8 & 12 12 48.4 & $-$60 43 15 & N/D \\
He 2-76        & 298.2$-$01.7  & 12 08.4 & $-$64 12 & ext \\
PHR1216-6039 & 298.6+01.9 & 12 16 09.1 & $-$60 39 32 & N/D \\
PHR1212-6407 & 298.6$-$01.5 & 12 12 28.0 & $-$64 07 11 & N/D \\
He 2-81        & 299.8$-$01.3  & 12 23.0 & $-$64 02 & ext \\
He 2-83        & 300.2+00.6  & 12 28.7 & $-$62 06 & ext \\
He 2-85        & 300.5$-$01.1  & 12 30.1 & $-$63 53 & ext \\
He 2-86        & 300.7$-$02.0  & 12 30.5 & $-$64 53 & ext \\
MPA1239-6048 & 301.5+02.0 & 12 39 46.3 & $-$60 48 17 & ext \\
PHR1244-6231 & 302.1+00.3 & 12 44 28.5 & $-$62 31 19 & N/D \\
PHR1246-6324 & 302.3$-$00.5 & 12 46 26.5 & $-$63 24 28 & ext \\
VBRC 4         & 302.6$-$00.9  & 12 48.5 & $-$63 50 & N/D \\
PHR1309-6457 & 304.7$-$02.1 & 13 09 01.7 & $-$64 57 18 & N/D \\
MPA1315-6338 & 305.6$-$00.9 & 13 15 30.4 & $-$63 38 43 & ext \\
Th 2-A         & 306.4$-$00.6  & 13 22.5 & $-$63 21 & N/D \\
BMP1322-6330 & 306.4$-$00.8 & 13 22 55.4 & $-$63 30 34 & N/D \\
MPA1324-6320 & 306.6$-$00.7 & 13 24 16.8 & $-$63 20 06 & N/D \\
MPA1330-6438 & 307.0$-$02.0 & 13 30 11.0 & $-$64 38 27 & N/D \\
PHR1326-6103 & 307.1+01.5 & 13 26 21.7 & $-$61 03 08 & N/D \\
PHR1332-6412 & 307.3$-$01.6 & 13 32 05.7 & $-$64 12 30 & N/D \\
MPA1337-6258 & 308.1$-$00.5 & 13 37 54.9 & $-$62 58 54 & ext \\
VBRC 5         & 309.2+01.3  & 13 44.0 & $-$60 50 & N/D \\
Vo 4           & 310.4+01.3  & 13 53.4 & $-$60 34 & ext \\
PHR1408-6106 & 312.1+00.3 & 14 08 51.7 & $-$61 06 27 & N/D \\
He 2-107       & 312.6$-$01.8  & 14 18.7 & $-$63 07 & ext \\
BMP1423-5923 & 314.4+01.3 & 14 23 59.7 & $-$59 23 38 & ext \\
PN 1417-5824   & 314.4+02.2  & 14 21.3 & $-$58 39 & ext \\
PHR1432-6138 & 314.5$-$01.0 & 14 32 05.0 & $-$61 38 42 & N/D \\
He 2-111       & 315.0$-$00.3  & 14 33.3 & $-$60 50 & ext \\
LoTr 9         & 315.7$-$01.2  & 14 41.3 & $-$61 20 & N/D \\
PHR1437-5949 & 315.9+00.3 & 14 37 53.2 & $-$59 49 25 & N/D \\
LoTr 10        & 316.3$-$01.3  & 14 46.3 & $-$61 14 & N/D \\
MPA1440-5802 & 316.9+01.8 & 14 40 27.4 & $-$58 02 19 & ext \\
He 2-114       & 318.3$-$02.0  & 15 04.2 & $-$60 54 & ext \\
PHR1457-5812 & 318.9+00.7 & 14 57 35.7 & $-$58 12 09 & ext \\
He 2-120       & 321.8+01.9  & 15 12.0 & $-$55 40 & ext \\
BMP1525-5823 & 322.0$-$01.3 & 15 25 59.4 & $-$58 23 02 & ext \\
MPA1525-5528 & 323.5+01.1 & 15 25 06.1 & $-$55 28 22 & N/D \\
MPA1539-5709 & 324.1$-$01.3 & 15 39 10.6 & $-$57 09 51 & N/D \\
PHR1548-5629 & 325.5$-$01.6 & 15 48 21.3 & $-$56 29 56 & N/D \\
FP1550-5639 & 325.6$-$01.8 & 15 50 04.6 & $-$56 39 13 & N/D \\
BMP1533-5319 & 325.7+02.2 & 15 33 20.6 & $-$53 19 01 & ext \\
MPA1541-5243 & 327.1+01.9 & 15 41 29.5 & $-$52 43 49 & ext \\
MPA1559-5552 & 327.1$-$02.0 & 15 59 11.1 & $-$55 52 18 & N/D \\
PHR1547-5214 & 328.0+01.8 & 15 47 09.6 & $-$52 14 44 & N/D \\
Sp 1           & 329.0+01.9  & 15 51.6 & $-$51 31 & ext \\
HeFa 1         & 329.5$-$02.2  & 16 12.5 & $-$54 24 & S/D \\
BMP1613-5406 & 329.8$-$02.1 & 16 13 02.0 & $-$54 06 32 & N/D \\
PHR1616-5324 & 330.7$-$02.0 & 16 16 54.2 & $-$53 24 40 & N/D \\
PHR1603-5016 & 331.3+01.6 & 16 03 53.7 & $-$50 16 52 & ext \\
He 2-145       & 331.4+00.5  & 16 09.0 & $-$51 02 & ext \\
Mz 3           & 331.7$-$01.0  & 16 17.2 & $-$51 59 & ext \\
PHR1626-5216 & 332.5$-$02.2 & 16 26 24.2 & $-$52 16 58 & N/D \\
He 2-152       & 333.4+01.1  & 16 15.3 & $-$49 13 & ext \\
PHR1619-4914 & 333.9+00.6 & 16 19 40.1 & $-$49 14 00 & ext \\
MmWe 1-6       & 334.3$-$01.4  & 16 31.1 & $-$50 26 & ext \\
He 2-169       & 335.4$-$01.1  & 16 34.3 & $-$49 21 & ext \\
PHR1637-4957 & 335.4$-$01.9 & 16 37 44.9 & $-$49 57 50 & ext \\
Pe 1-6         & 336.2+01.9  & 16 23.9 & $-$46 42 & ext \\
BMP1636-4529 & 338.6+01.1 & 16 36 58.7 & $-$45 29 29 & N/D \\
MPA1657-4447 & 341.5$-$01.1 & 16 57 13.3 & $-$44 47 18 & N/D \\
PHR1702-4443 & 342.0$-$01.7 & 17 02 04.3 & $-$44 43 20 & N/D \\
H 1-3          & 342.7+00.7  & 16 53.5 & $-$42 39 & N/D \\
He 2-198       & 342.9$-$02.0  & 17 06.4 & $-$44 13 & ext \\
HtTr 5         & 343.3$-$00.6  & 17 01.5 & $-$43 06 & N/D \\
PHR1653-4143 & 343.5+01.2 & 16 53 55.3 & $-$41 44 00 & N/D \\
H 1-5          & 343.9+00.8  & 16 57.4 & $-$41 38 & ext \\
H 1-6          & 344.2$-$01.2  & 17 07.0 & $-$42 41 & ext \\
H 1-7          & 345.2$-$01.2  & 17 10.4 & $-$41 53 & ext \\
IC 4637        & 345.4+00.1  & 17 05.2 & $-$40 53 & ext \\
MPA1715-3903 & 348.0$-$00.3 & 17 15 16.1 & $-$39 03 48 & N/D \\
PHR1710-3732 & 348.7+01.3 & 17 10 12.3 & $-$37 32 51 & N/D \\
PHR1724-3859 & 349.1$-$01.7 & 17 24 30.7 & $-$38 59 44 & N/D \\
NGC 6337       & 349.3$-$01.1  & 17 22.2 & $-$38 29 & ext \\
NGC 6302       & 349.5+01.0  & 17 13.8 & $-$37 06 & ext \\
MPA1727-3851 & 349.6$-$02.1 & 17 27 36.8 & $-$38 51 09 & N/D \\
H 1-26         & 350.1$-$03.9 & 17 36 29.90  & $-$39 21 56.8  & ext \\
H 1-28         & 350.5$-$05.0 & 17 42 54.066  & $-$39 36 24.04  & ext \\
PPA1734-3817 & 350.8$-$03.0 & 17 34 52.8  & $-$38 17 19 & N/D \\
SB 33          & 351.2$-$06.3  & 17 50 27.700  & $-$39 40 17.40  & N/D \\
SB 34          & 351.5$-$06.5  & 17 52 09.386  & $-$39 32 14.52  & S/D \\
H 1-37         & 351.6$-$06.2 & 17 50 44.569  & $-$39 17 25.98  & ext \\
SB 35          & 351.7$-$06.6  & 17 53 02.87  & $-$39 24 08.9 & S/D \\
SB 36          & 352.0$-$06.7  & 17 54 20.83  & $-$39 10 37.9 & S/D \\
H 1-12         & 352.6+00.1 & 17 26 24.258  & $-$35 01 41.88 & ext \\
H 1-13         & 352.8$-$00.2 & 17 28 27.46  & $-$35 07 31.8  & ext \\
Fg 3           & 352.9$-$07.5 & 18 00 11.82  & $-$38 49 52.6 & S/D \\
H 1-38         & 353.2$-$05.2 & 17 50 45.216  & $-$37 23 54.18 & N/D \\
PPA1738-3546 & 353.3$-$02.2 & 17 38 16.2  & $-$35 46 27 & N/D \\
PPA1740-3551 & 353.5$-$02.6 & 17 40 21.4  & $-$35 51 31  & N/D \\
JaFu 2         & 353.5$-$05.0  & 17 50 11.1  & $-$37 03 27  & N/D \\
H 1-52         & 354.4$-$07.8 & 18 04 57.594  & $-$37 38 07.92  & S/D \\
PHR1722-3210 & 354.5+02.4 & 17 22 11.7  & $-$32 10 45  & N/D \\
PPA1740-3437 & 354.5$-$02.0a & 17 40 30.5  & $-$34 37 17  & N/D \\
Ta 139         & 354.6+04.9  & 17 12 53.539  & $-$30 40 05.70  & N/D \\
SB 40          & 354.7$-$07.2  & 18 02 55.69  & $-$37 08 13.1  & S/D \\
H 1-31         & 355.1$-$02.9 & 17 45 32.104  & $-$34 33 55.32  & N/D \\
M 3-21         & 355.1$-$06.9 & 18 02 32.33  & $-$36 39 11.3 & S/D \\
SB 42          & 355.3$-$07.5  & 18 05 52.556  & $-$36 45 37.40  & S/D \\
Ta 138         & 355.4+02.3  & 17 25 03.470  & $-$31 28 38.50  & ext \\
PHR1740-3324 & 355.6$-$01.4 & 17 40 54.9  & $-$33 24 18 & N/D \\
PHR1721-3027 & 355.8+03.5 & 17 21 31.7  & $-$30 27 20  & N/D \\
SB 43          & 355.8$-$08.7  & 18 12 23.684  & $-$36 52 53.72  & N/D \\
PPA1741-3302 & 356.0$-$01.4 & 17 41 33.4  & $-$33 02 15  & N/D \\
SB 44          & 356.0$-$07.4A & 18 07 07.953  & $-$36 02 45.64  & S/D \\
SB 45          & 356.0$-$07.4B & 18 06 52.495  & $-$36 06 42.82  & S/D \\
PPA1747-3341 & 356.1$-$02.7 & 17 47 04.8  & $-$33 41 03  & N/D \\
SB 46          & 356.1$-$08.6  & 18 12 39.873  & $-$36 31 49.16  & N/D \\
PPA1751-3401 & 356.2$-$03.6 & 17 51 06.9  & $-$34 01 41  & N/D \\
M 3-49         & 356.3$-$06.2 & 18 02 32.06  & $-$35 13 14.3  & S/D \\
SB 47          & 356.3$-$07.3  & 18 07 21.472  & $-$35 45 44.04  & N/D \\
SB 48          & 356.4$-$06.8  & 18 05 14.396  & $-$35 28 07.54  & S/D \\
H 1-57         & 356.6$-$07.8 & 18 09 49.240  & $-$35 44 13.32 & S/D \\
H 1-51         & 356.7$-$06.4 & 18 04 29.304  & $-$34 58 00.98 & S/D \\
M 4-4          & 357.0+02.4 & 17 28 50.29  & $-$30 07 45.1 & N/D \\
Th 3-24        & 357.1+01.9 & 17 30 51.354  & $-$30 17 12.49  & N/D \\
TaJu 18        & 357.1+04.4 & 17 21 37.90  & $-$28 55 13.9 & N/D \\
M 3-50         & 357.1$-$06.1 & 18 04 05.183  & $-$34 28 37.39  & S/D \\
H 1-42         & 357.2$-$04.5 & 17 57 25.17  & $-$33 35 42.9  & ext \\
H 2-7          & 357.3+04.0 & 17 23 24.936  & $-$28 59 06.05  & ext \\
SB 50          & 357.3$-$06.5  & 18 06 08.301  & $-$34 33 30.72  & S/D \\
PPA1749-3216 & 357.5$-$02.4 & 17 49 37.82  & $-$32 16 27.9  & N/D \\
PPA1802-3350 & 357.5$-$05.5 & 18 02 23.8  & $-$33 50 48  & N/D \\
Pe 1-11        & 358.0$-$05.1 & 18 01 42.89  & $-$33 15 28.3  & N/D \\
M 3-8          & 358.2+04.2 & 17 24 52.152  & $-$28 05 54.61  & ext \\
Bl D           & 358.2$-$01.1 & 17 46 02.818  & $-$31 03 39.33  & N/D \\
SB 52          & 358.3$-$07.3  & 18 11 39.900  & $-$34 00 21.99  & N/D \\
JaSt 3         & 358.4+01.6  & 17 35 22.745  & $-$29 22 17.40 & N/D \\
NGC 6563       & 358.5$-$07.3 & 18 12 02.753  & $-$33 52 07.14  & ext \\
JaSt 4         & 358.6+01.7  & 17 35 37.47  & $-$29 13 17.7 & N/D \\
JaSt 58 & 358.6$-$01.1  & 17 46 52.20  & $-$30 37 42.8 & N/D \\
PHR1752-3116 & 358.7$-$02.5 & 17 52 36.5  & $-$31 16 27  & N/D \\
SB 53          & 358.7$-$05.1  & 18 03 28.532  & $-$32 37 26.46  & N/D \\
H 1-50         & 358.7$-$05.2 & 18 03 53.461  & $-$32 41 42.15  & N/D \\
JaSt 5         & 358.8+01.7  & 17 35 52.65  & $-$28 58 25.7 & N/D \\
Th 3-26        & 358.8+03.0 & 17 31 09.298  & $-$28 14 50.43  & ext \\
JaSt 9         & 359.0+01.1  & 17 38 45.91  & $-$29 09 01.6 & N/D \\
M 3-48         & 359.0$-$04.1 & 17 59 56.823  & $-$31 54 27.46  & ext \\
M 1-29         & 359.1$-$01.7 & 17 50 18.003  & $-$30 34 54.90  & ext \\
H 2-33         & 359.4$-$03.4 & 17 58 12.535  & $-$31 08 05.99 & ext \\
KnFs 3         & 359.7$-$04.4 & 18 02 52.931  & $-$31 23 58.49  & N/D \\
PPA1738-2800 & 359.9+01.8 & 17 38 11.8  & $-$28 00 07  & N/D \\
\end{longtable}
\end{landscape}
}% End \longtabL

%--------------APPENDIX----------------------------------------------------

\newpage
\appendix
%\section{Appendix A \\
\section{Discussion about aperture}\label{apendi}

As we are dealing with extended sources, some special care had to be taken.
Thus, we characterised the object sizes by means the 
Petrosian radius \citep{1976ApJ...209L...1P, 2001AJ....122.1104Y} provided by CASU
(see Table~\ref{tablita-3}).
However, the relation between a measured size and the real size of an extended
nebula depends on the resolution/seeing versus actual size, and on the
assumed geometry. There is a discussion on this in \cite{2000MNRAS.314...99V}.

\begin{figure}
   \centering
   \includegraphics[width=8cm]{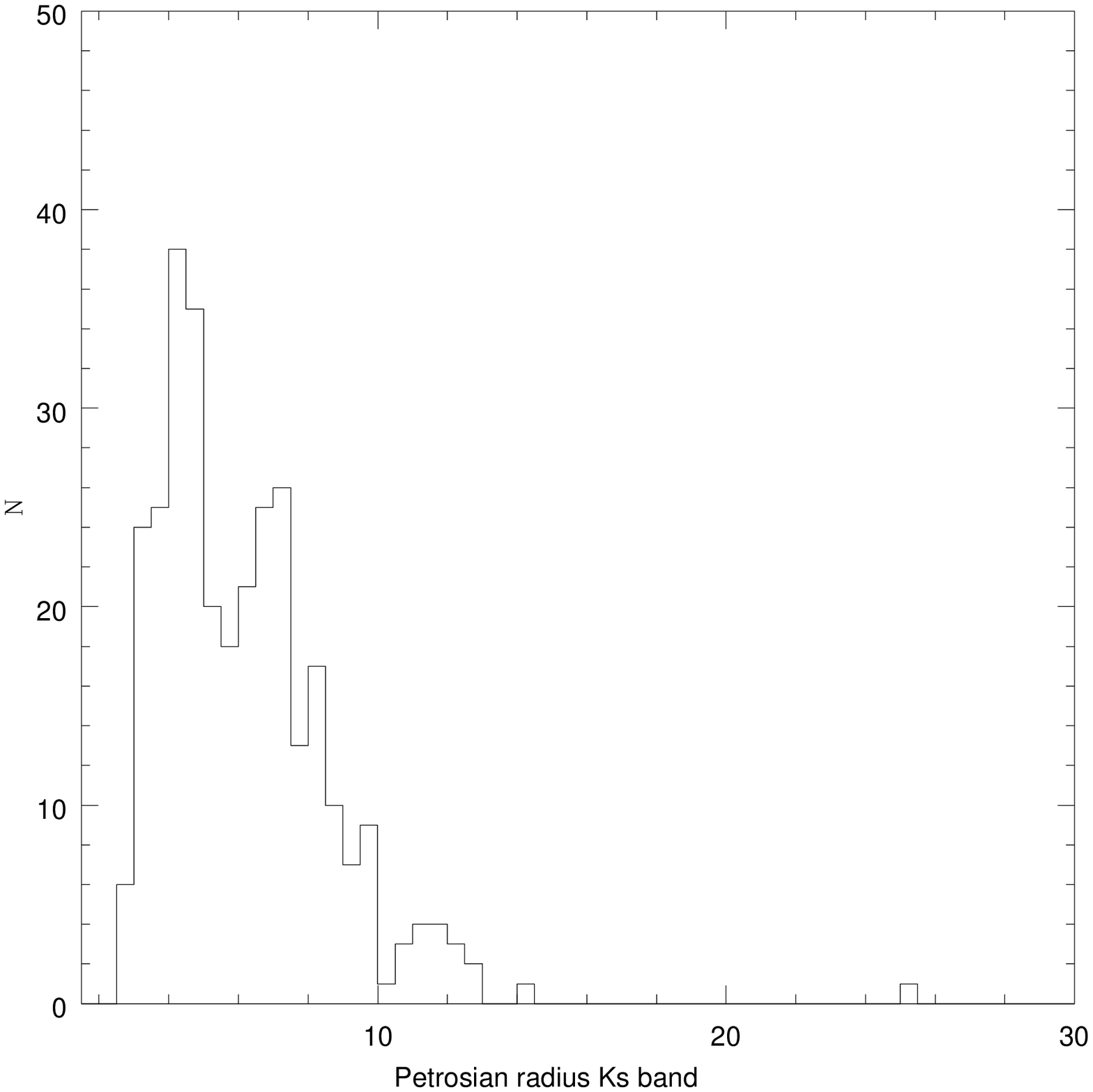}
   \includegraphics[width=8cm]{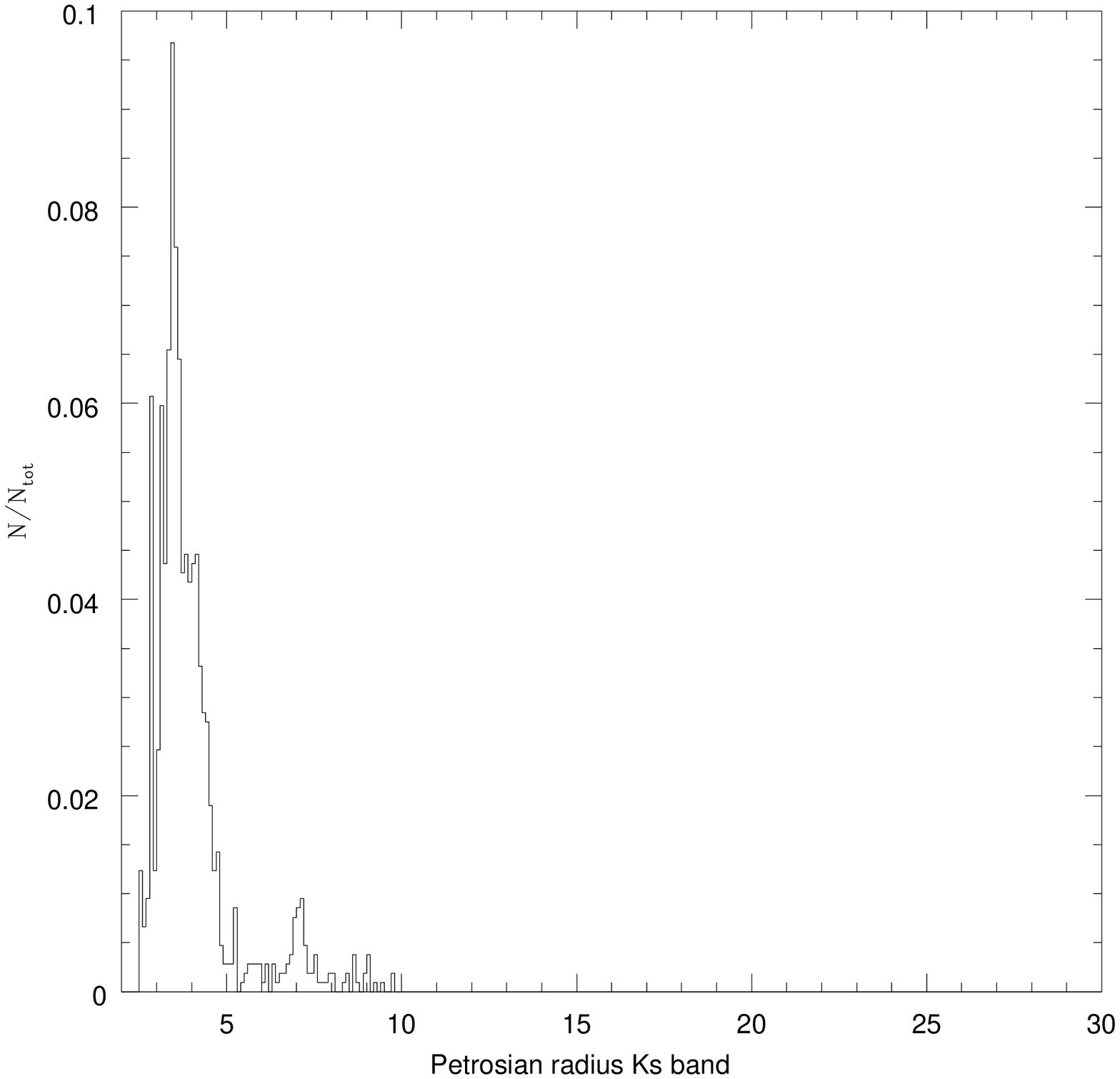}
      \caption{Distribution of Petrosian radii, in pixels.
              Upper panel, distribution for our sample of PNe (data from table~\ref{tablita-3}).
              Lower panel, distribution for a sample of sources clasiffied like stars by CASU.}
         \label{extra2}
   \end{figure}

According to Fig.~\ref{extra2} (upper panel), roughly 80\% of
PNe in our sample have a Petrosian radius less than 8 px,
equivalent to $2.6\arcsec$. In this sense, the magnitudes 
for our sample of PNe should be those of  
aperture-6 (= $2.83\arcsec$) in the CASU catalogue.

\begin{figure}
   \centering
   \includegraphics[width=8cm]{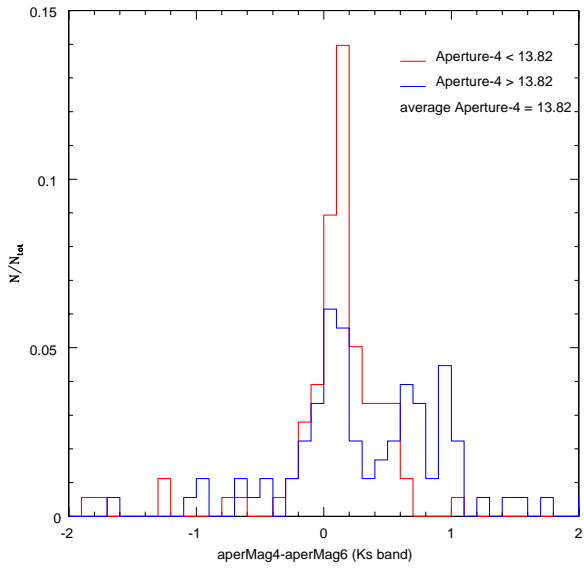}
      \caption{Comparison between magnitudes with aperture-6 ($2.83\arcsec$) and aperture-4 
               ($1.41\arcsec$) for our sample of PNe.}
         \label{a4a6}
   \end{figure}

\begin{figure}
   \centering
   \includegraphics[width=6cm]{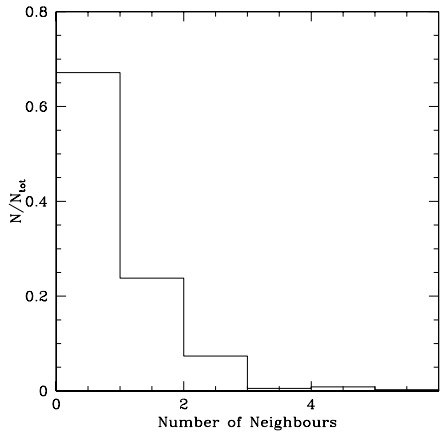}
      \caption{Distribution of number of contaminant stars around each PNe of our sample.}
         \label{contamina}
   \end{figure}

In order to evaluate how important the differences 
between magnitudes obtained with an aperture 4 and 6 are,
we assemble a histogram of $aperMag4 - aperMag6$ (Fig.~\ref{a4a6}).
Clearly, differences between both magnitudes are important.
The tail towards brighter aperture-6 magnitudes may indicate the effect
of extended sources and 
the presence of neighbour stars within the aperture. 
To evaluate this later hypothesis, we computed the number of stars
that are closer than $2.6\arcsec$ to each PNe. 
Surprisingly, about 70\% of the sample do not have any contaminant
star (Fig.~\ref{contamina}),
pointing out that the differences should be due to other effects.

Alternatively, we analysed the sky subtraction performed by the CASU pipeline 
and found that the error introduced by this task is very significant. 
The weakest sources, as the PNe in our sample, are more 
affected by sky subtraction (see Fig.~\ref{a4a6}).  
Also, we note that the magnitudes obtained with aperture-4 compare better
with those of 2MASS (Fig.~\ref{j2-jv})
than the aperture-6 ones,
 and the distribution of PNe in the \textit{$J-H$} Vs. 
\textit{$H-K_{\rm s}$} diagram is more scattered.

Although it seems more appropriate to obtain the magnitudes with an 
aperture of $2.83\arcsec$, increasing aperture implies increasing 
uncertainty in the magnitudes (due the sky subtraction), and the magnitudes obtained
with larger aperture are affected by larger errors. 
In this sense, we demonstrate that the more accurate magnitudes are
those computed with the CASU pipeline using the aperture of $1.41\arcsec$.

\end{document}